\documentstyle[12pt,epsfig]{article}

\begin{document}

\begin{titlepage}

\rightline{August 2010}

\vskip 2cm

\centerline{\large \bf  
A comprehensive analysis of the dark matter direct} 

\vskip 0.2cm
\centerline{
\large \bf detection experiments in the mirror dark matter framework}

\

\vskip 2.2cm

\centerline{R. Foot\footnote{
E-mail address: rfoot@unimelb.edu.au}}

\vskip 0.7cm

\centerline{\it School of Physics,}
\centerline{\it University of Melbourne,}
\centerline{\it Victoria 3010 Australia}

\vskip 2cm

\noindent

Mirror dark matter offers a framework to explain the existing dark
matter direct detection experiments.
%including the impressive DAMA annual modulation signal, CoGeNT low energy excess 
%and hints from CDMS electron scattering, CDMS/Ge, Edelweiss and CRESSTII.  
Here we confront this theory with the most
recent experimental data, paying attention to the various known
systematic uncertainties,
in quenching factor, detector resolution, galactic rotational velocity
and
velocity dispersion. We perform a detailed analysis of the DAMA and CoGeNT experiments 
assuming a negligible channeling fraction and find that the data
can be fully explained within the mirror dark matter
framework. We also show that the mirror dark matter candidate can explain recent data from the CDMS/Ge, 
EdelweissII and CRESSTII experiments and we point out ways in which the theory can be further tested
in the near future.

\end{titlepage}

\section{Introduction}

The field of dark matter direct detection has blossomed in recent times,
with exciting positive signals from DAMA\cite{dama,dama2,dama3},
CoGeNT\cite{cogent}, as well as interesting
hints from CDMS/Ge\cite{cdmsge} and CDMS electron
scattering\cite{cdmselectron}.
Very recently, more exciting evidence for the direct detection of dark matter
has arisen from the CRESSTII\cite{cresst} and EdelweissII\cite{edelweiss} experiments.
 
Mirror dark matter has emerged as a simple predictive framework which can explain all of the
direct detection experiments\cite{mm,mm2,mmelectron,mmcdms,mmcogent}. 
The purpose
of this article is to provide a comprehensive update of the experimental status 
of the mirror dark matter candidate, paying particular attention to the various known systematic
uncertainties,in quenching factor, detector resolution, galactic rotational velocity
and velocity dispersion. 

%We also wish to update this interpretation
%of the direct detection experiments in the light of the recent study of
%ref.\cite{gelmini}
%where they concluded that the channeling fraction in $NaI$ crystals is
%likely to be very small 
%($< 1\%$ channeling for both $Na$ and $I$).
%We will show that even with negligible channeling fraction, the DAMA
%experiment can be fully explained within the mirror dark matter
%framework. Importantly this
%explanation remains consistent with the positive results of CoGeNT and
%the null results of the other experiments.

Recall, mirror dark matter posits that the inferred dark matter in the
Universe arises from
a hidden sector which is an exact copy of the standard model
sector\cite{flv} (for a review
see ref.\cite{review}).
That is, 
a spectrum of dark matter particles of known masses are predicted: $e',
H', He', O', Fe',...$ (with
$m_{e'} = m_e, m_{H'} = m_H,$ etc). 
The galactic halo is then presumed to be composed predominately of a
spherically distributed 
self interacting mirror particle plasma comprising these
particles\cite{sph}. 
In addition to gravity, ordinary and mirror particles
interact with each other via (renormalizable) 
photon-mirror photon kinetic mixing\cite{flv,he}:
\begin{eqnarray}
{\cal L}_{mix} = \frac{\epsilon}{2} F^{\mu \nu} F'_{\mu \nu}\ ,
\end{eqnarray}
where $F_{\mu \nu}$ ($F'_{\mu \nu}$) is the ordinary (mirror) $U(1)$
gauge boson 
field strength tensor. This interaction enables mirror charged particles
to couple to
ordinary photons with electric charge $q = \epsilon e$ and thus 
allows mirror particles to elastically scatter off ordinary particles.
This means that mirror dark matter 
can be probed in dark matter direct detection experiments.
It turns out that this simple predictive theory can explain the DAMA annual
modulation
signal, the CoGeNT low energy excess as well as hints from CDMS, Edelweiss and CRESSTII
consistently
with the null results of other experiments.

The outline of this paper is as follows.
In section 2 we provide a brief review of the mirror dark matter theory.
In section 3 we provide some necessary technical details: cross section
and halo
distribution which are characteristic of mirror dark matter.
% and quite different to the standard WIMP model. 
In section 4 (5), we examine the implications of the most recent DAMA
(CoGeNT)
data for the mirror dark matter theory. 
These experiments are sensitive to dark matter particles heavier than
around
10 GeV which makes them excellent probes of the dominant mirror metal component
of the galactic
halo, $A'$.  We show that these experiments
can be simultaneously explained and lead to a measurement of the
parameters:
$\epsilon \sqrt{\xi_{A'}}$ and $m_{A'}$ both of which are consistent
with the theoretical
expectations of $\epsilon \sim 10^{-9}$ (from galactic halo energy
balance) and 
$A' \sim O' \Rightarrow m_{A'} \sim 16m_p$ (from analogy with the 
ordinary matter sector).
%[We expect the mirror metal spectrum to be dominated by mirror oxygen,
%for much
%the same reason that oxygen is the dominant heavy element in the
%ordinary matter sector; 
%which is the exceptional stability of the oxygen nucleus].
In section 5 we also show that the DAMA and CoGeNT signals are 
consistent with the results of the other experiments including the null
results of XENON100 and CDMS/Si.
In section 6 we examine the constraints on $e'$ scattering from the DAMA
absolute rate.
% as well as the hints of $e'$ detection from the CDMS electron
% scattering data. 
We show that these constraints
%these experiments 
when combined with the DAMA and CoGeNT
data suggest a halo mirror metal proportion $\xi_{A'} \stackrel{>}{\sim}
10^{-2}$. 
In section 7 we examine recent data from the CDMS/Ge, EdelweissII and CRESSTII experiments.
CDMS/Ge and Edelweiss are excellent probes of the anticipated $Fe'$ component, and the data
are consistent with a $Fe'$ component with mass fraction: 
$\xi_{Fe'}/\xi_{A'} \sim 10^{-2}$. We also point out that the CRESSTII experiment is potentially
sensitive to both $A'$ and $Fe'$ components and their recently announced low energy excess
can be explained by $A'$ and $Fe'$ interactions.
%We speculate that a rapid period of mirror star formation and evolution
%in the early
%stages of the universe might have depleted the $He'$.
In section 8 we draw our conclusions.

\section{A brief review of mirror dark matter}

Mirror dark matter conjectures that the inferred dark matter in the
Universe arises from
a hidden sector which is an exact copy of the standard model sector.
That is, the standard model of particle physics is extended:
\begin{eqnarray}
{\cal L} = {\cal L}_{SM} (e, u, d, \gamma,...) + {\cal L}_{SM} (e', u',
d', \gamma', ...)\ .
\end{eqnarray}
Such a theory can be theoretically well motivated from symmetry
considerations 
if left and right handed chiral
fields are interchanged in the extra sector. In this way space-time
parity symmetry and in fact
the full Poincar$\acute{e}$ group
can be realized as an unbroken symmetry of nature, and for this reason
we refer to the particles in
the extra sector as mirror particles. The standard model extended with 
a mirror sector was first studied in ref.\cite{flv} and shown to be 
a phenomenologically consistent renormalizable extension
of the standard model. The concept, though, has a long history dating back prior
to the advent of the
standard model of particle interactions\cite{lee}. 
For a review and more complete list of references see ref.\cite{review}.

If we include all interaction terms consistent with renormalizability
and the symmetries of
the theory then we must add to the Lagrangian a $U(1)$ kinetic mixing
interaction\cite{he,flv} and Higgs - mirror Higgs quartic
coupling\cite{flv}:
\begin{eqnarray}
{\cal L}_{mix} = \frac{\epsilon}{2} F^{\mu \nu} F'_{\mu \nu} + 
\lambda \phi^{\dagger}\phi \phi'^{\dagger}\phi' \ ,
\label{mix}
\end{eqnarray}
where $F_{\mu \nu}$ ($F'_{\mu \nu}$) is the ordinary (mirror) $U(1)$
gauge boson 
field strength tensor and
$\phi$ ($\phi'$) is the electroweak Higgs (mirror Higgs) field.
The most general Higgs potential, including the quartic Higgs mixing
term (above)
was studied in ref.\cite{flv} and shown to have 
the vacuum $\langle \phi \rangle = \langle \phi' \rangle$ for
a large range of parameters. With this vacuum,
the mirror symmetry is unbroken and consequently the masses
of the mirror particles are all identical to their ordinary
matter counterparts. 

In this framework, dark matter is comprised of a spectrum of stable
massive mirror particles: 
$e', H', He', O'...$ etc, with masses $m_{e'} = m_e, m_{H'} = m_H,$
$m_{He'} = m_{He}$ etc.  
To explain the rotation curves in spiral galaxies, the dark matter needs
to be
roughly spherically distributed in galactic halos. Given the upper limit
on compact star
sized objects (MACHOs) in the halo from microlensing observations, 
roughly $f_{macho} \stackrel{<}{\sim} 0.2-0.3$
depending on the assumptions\cite{macholimits}, we then expect the
mirror particles to be 
distributed predominately
as a hot gaseous spherical halo surrounding the collapsed disk of ordinary
matter\cite{sph}
\footnote{
Naturally a MACHO subcomponent consisting of mirror white dwarfs, mirror
neutron stars etc are also
expected and can be probed by microlensing observations. Since most of
the stellar mass is ejected as gas
in the explosions producing these stellar remnants, it is plausible that
the MACHO mass fraction
can satisfy the observational limit of $f_{macho} \stackrel{<}{\sim}
0.2-0.3$.
}.

Observations of colliding clusters, such as the bullet
cluster\cite{bulletcluster} indicate that dark matter
does not have self interactions on galaxy cluster scales. This suggests
that the gaseous
mirror dark matter component is confined to galactic halos (c.f.
\cite{silagadze}). 
Gravity and the  mirror particle self 
interactions may well be sufficient to achieve this.

A dissipative dark matter candidate like mirror matter can only survive
in an extended spherical
distribution in galaxies without collapsing
if there is a substantial heating mechanism to replace the energy lost
due to radiative cooling.
In fact, ordinary supernova
can plausibly supply the required heating if the photon and mirror
photon
are kinetically mixed with $\epsilon \sim 10^{-9}$\cite{sph}
\footnote{ 
A mirror sector with such kinetic mixing is consistent with
all known laboratory, astrophysical and 
cosmological constraints\cite{lab}.}.
For kinetic mixing of this magnitude 
about half of the total energy emitted in ordinary Type II Supernova
explosions ($\sim 3\times 10^{53}$ erg) will be in the form of light
mirror 
particles ($\nu'_{e,\mu,\tau}$, $e'^{\pm}, \gamma'$)
originating from kinetic mixing induced plasmon decay into $e'^+ e'^-$
in the 
supernova core\cite{rafelt}. Given the observed rate of Supernova's in
our galaxy of about 1 per century,
this implies a heating of the halo
(principally due to the $e'^{\pm}$ component),
of around:
\begin{eqnarray}
L^{SN}_{heat-in} \sim \frac{1}{2} \times 3\times 10^{53}\ erg 
{1 \over 100\ years} \sim 10^{44}\ {\rm erg/s, \ for \ the \ Milky\ Way\ .
}
\label{4z}
\end{eqnarray}
It turns out that this matches (to within uncertainties) the energy lost
from the halo due to
radiative cooling\cite{sph}:
\begin{eqnarray}
L^{halo}_{energy-out} = \Lambda \int n^2_{e'} 4\pi r^2 dr \sim
10^{44} \ {\rm erg/s, \ for
\ the \ Milky \ Way}.
\label{5z}
\end{eqnarray}
In other words, a gaseous mirror particle halo can potentially survive
without collapsing because
the energy lost due to dissipative interactions can be replaced by the
energy from ordinary supernova
explosions. Presumably there are feedback mechanisms which
maintain this balance. For example if $L^{halo}_{energy-out} > L^{SN}_{heat-in}$
then the halo would contract which in turns increases the gravitational pull on the
ordinary matter component. This compression of the ordinary matter 
component should increase ordinary star formation rates, thereby increasing $L^{SN}_{heat-in}$
until the energy is balanced.
In this way the ordinary supernova rate might be dynamically adjusted
so that the halo is stabilized.
Extending these ideas to galaxies beyond the Milky Way,
the hypothesized connection between Supernova rates and dark matter
distribution might ultimately provide a dynamical justification for the 
empirical Tully-Fischer and Faber-Jackson relations.

The spherically distributed mirror particle plasma is likely to be
far too hot for much mirror star formation to occur at the present epoch. 
Thus, we do not expect significant heating of the ordinary matter
sector from mirror Supernovas at the present time.
However during the first billion years or so
the situation might have been the reverse. That is, the early stages of galaxies may
have witnessed rapid mirror star formation and evolution 
and little ordinary star formation, due ultimately to the effects
of asymmetric initial conditions in the Early Universe. In particular,
for $T' \ll T$ in the early Universe (required to achieve successful big
bang nucleosynthesis and large scale structure formation)\footnote{
See ref.\cite{some} for further discussions about early Universe cosmology with mirror dark matter.}
and $\epsilon \sim 10^{-9}$ the primordial mirror $He'$ abundance is 
expected to be relatively high, $Y_{He'} \approx 0.9$\cite{bbn}. With such initial conditions
the evolution rate of stars is dramatically increased by several orders of magnitude\cite{starevolution}.
In other words, we surmise that the required asymmetric evolution of the ordinary
and mirror matter components
in galaxies, originates in the complex interactions between the ordinary
and mirror components 
(such as the energy transfer to the mirror sector from ordinary
supernova's) 
together with asymmetric initial conditions in the early Universe.

Since mirror charged particles have an electric charge induced via the
kinetic mixing, 
mirror nuclei can 
interact with ordinary nuclei via spin independent Rutherford elastic
scattering.
It turns out that the positive results of DAMA and CoGeNT can be
explained by
such scattering from a putative $\sim O'$ component if $\epsilon \sim
10^{-9}$. This
provides important experimental evidence in favour of the mirror dark
matter candidate, which we
now examine in detail.

\section{Interaction cross-section and galactic mirror dark matter distribution}

The interaction rate in an experiment depends on the cross-section,
$d\sigma/dE_R$, 
and halo velocity distribution, $f(v)$.
The photon-mirror photon kinetic mixing enables a mirror nucleus [with
mass and atomic
numbers $A',\ Z'$ and velocity $v$] to elastically scatter
with an ordinary nucleus [presumed at rest with mass and atomic numbers
$A,\ Z$].
In fact the cross-section is just of the standard Rutherford form
corresponding
to a particle of electric charge $Ze$ scattering off a particle of
electric charge
$\epsilon Z'e$. The cross-section can be expressed in terms of the 
recoil energy of the ordinary nucleus, $E_R$\cite{mm}:
\begin{eqnarray}
{d\sigma \over dE_R} = {\lambda \over E_R^2 v^2}\ ,
\label{cs}
\end{eqnarray}
where 
\begin{eqnarray}
\lambda \equiv {2\pi \epsilon^2 Z^2 Z'^2 \alpha^2 \over m_A} F^2_A
(qr_A) F^2_{A'} (qr_{A'}) \ ,
\end{eqnarray}
and $F_X (qr_X)$ ($X = A, A'$) are the form factors which
take into account the finite size of the nuclei and mirror nuclei.
[The quantity $q = (2m_A E_R)^{1/2}$ is the momentum transfer and $r_X$
is the effective
nuclear radius]\footnote{
Unless otherwise specified,
we use natural units, $\hbar = c = 1$ throughout.}.
A simple analytic expression for
the form factor, which we adopt in our numerical work, is the one
given by Helm\cite{helm,smith}:
\begin{eqnarray}
F_X (qr_X) = 3{j_1 (qr_X) \over qr_X} e^{-(qs)^2/2}\ ,
\end{eqnarray}
with $r_X = 1.14 X^{1/3}$ fm, $s = 0.9$ fm and $j_1$ is the spherical
Bessel function of index 1.

The halo mirror particles are presumed to form a self interacting 
plasma at an isothermal temperature $T$.
This means that the halo distribution function is given by a 
Maxwellian distribution:
\begin{eqnarray}
f_i (v) &=& e^{-\frac{1}{2} m_i v^2/T}  \nonumber \\  
 &=& e^{-v^2/v_0^2[i]}\ , 
\end{eqnarray}
where the index $i$ labels the particle type [$i=e', H', He', O',
Fe',...$] and $v_0^2[i] \equiv 2T/m_i$.
The dynamics of such a mirror particle plasma has been investigated
previously\cite{sph,mm},
where it was found that the condition of hydrostatic equilibrium implied
that the
temperature of the plasma satisfied:
\begin{eqnarray}
T \simeq  {1 \over 2} \bar m v_{rot}^2 \ ,
\label{4}
\end{eqnarray}
where $\bar m = \sum n_i m_i/\sum n_i$ is the mean mass of  
the particles in the plasma, and $v_{rot} \approx 254\pm 16$ km/s is the
galactic rotational velocity of the Milky Way\cite{rot}.
The velocity dispersion of the particles in the mirror particle halo
evidently depends on the particular particle species and satisfies:
\begin{eqnarray}
v_0^2 [i] = v_{rot}^2 \frac{\overline{m}}{m_i} \ .
\label{dis}
\end{eqnarray}
Note that if $m_i \gg \overline{m}$, then $v_0^2[i] \ll v_{rot}^2$. 
Consequently  heavy
mirror nuclei have their velocities (and hence energies) relative to the
Earth
boosted by the Earth's (mean) rotational velocity
around the galactic center, $\approx v_{rot}$. 
This allows a mirror nuclei in the	`oxygen' mass range $\sim m_0
\approx 15$ GeV 
to provide a significant annual modulation signal in the
energy region probed by DAMA ($E_R > 2$ keVee) which, as we will
see,
has the right properties to fully account for the data presented
by the DAMA collaboration\cite{dama,dama2,dama3}.

According to the above considerations, in order to calculate the
velocity dispersion, $v_0^2[i]$,
we need to estimate the mean mass of the mirror particles in the plasma. 
The plasma is expected to be completely ionized since it turns out that
the temperature of the plasma is $T \approx \frac{1}{2}$ keV. We start
by making the simplifying assumption that the
mirror metal component of the plasma is dominated by a single element,
$A'$.
%[In the ordinary matter case, this assumption would be valid, with $A =$
%Oxygen].
Under this assumption, the plasma consists of 
$e', H', He'$ and $A'$. It is straightforward to estimate $\bar m$: 
\begin{eqnarray}
{\bar m \over m_p} \simeq {1 \over 2 - \frac{5}{4} \xi_{He'} + 
\xi_{A'} ( {1 \over A'} - \frac{3}{2})}\ , 
\label{mary}
\end{eqnarray}
where $\xi_{i} \equiv {n_{i} m_{i} \over n_{H'} m_{p} + n_{He'}m_{He} + 
n_{A'} m_{A'} } $ is the halo mass fraction of species $i$, $m_p$ is the
proton mass and $A'$ is the mass number. 
Thus, combining Eq.(\ref{dis}) and Eq.(\ref{mary}) we have
\begin{eqnarray}
v^2_0 [A'] =
{v_{rot}^2  \over A'[2 - \frac{5}{4} \xi_{He'} + 
\xi_{A'} ( {1 \over A'} - \frac{3}{2})]}\ .
\label{mary2}
\end{eqnarray}
If we vary $\xi_{A'}$ between $0$ and $1$, then we obtain a lower and upper limit
for $v_0^2 [A']$:
\begin{eqnarray}
 {1 \over A'(2 - \frac{5}{4} \xi_{He'})}  < {v_0^2[A'] \over v_{rot}^2}
< {1 \over 1 + \frac{A'}{2}}\ .
\label{limits2}
\end{eqnarray}
Mirror BBN studies\cite{bbn} indicate that the primordial Helium mass
fraction, $Y_{He'}$, 
is relatively high, with $Y_{He'} \simeq 0.9$, which is quite unlike the
case of ordinary matter.
However, like the ordinary matter sector, the primordial value 
for $\xi_{A'}$ is expected\cite{bbn} to be small $\xi_{A'} \ll 1$. That
is, heavy mirror elements are
anticipated to be synthesised in mirror stars.  If the net $A'$
production from mirror stars
remains subdominant, i.e. $\xi_{A'} \ll 1$, then we expect that $v_0 [A']$ to be 
given by the lower limit in Eq.(\ref{limits2}) with
$\xi_{He'} = Y_{He'} \simeq 0.9$. 
The situation where $\xi_{A'} \approx 1$ corresponds to extremely efficient mirror star
formation and evolution, and is a priori possible. 
We will therefore consider both limiting cases for $v_0 [A']$.
It turns out, though, that our 
results are relatively insensitive to the possible variation of $v_0 [A']$ given by Eq.(\ref{limits2})
simply because $v_0 [A']/v_{rot} << 1$.

\section{The DAMA experiment}

The DAMA experiments\cite{dama,dama2,dama3} employ large mass
scintillation sodium iodide detectors operating
in the Gran Sasso National Laboratory. These experiments
were initially operating with a target mass of around 100 kg and since
2003 with a target of $\sim 250$ kg.
These experiments have been running for more than 13 years and now have a
cumulative exposure of $1.17\ {\rm ton}\times {\rm year}$.
Importantly, the DAMA experiments have consistently observed a 
positive dark matter signal, with statistical significance of around 8.9
$\sigma$ C.L.\cite{dama3}.

The DAMA experiments utilize the annual modulation signature, which
provides a ``smoking
gun'' signal for dark matter. The idea\cite{idea} is very simple.
The interaction rate must vary periodically since it depends on the
Earth's velocity, $v_E$, which 
modulates due to the Earth's motion around the Sun. That is,
\begin{eqnarray}
R(v_E) = R (v_{\odot}) + \left( \frac{\partial R}{ \partial
v_E}\right)_{v_{\odot}} \Delta v_E \cos \omega
(t-t_0)
\end{eqnarray}
where $v_{\odot} = v_{rot} + 12$ km/s is the sun's velocity with respect
to the galactic halo and
$\Delta v_E \simeq 15$ km/s, $\omega \equiv  2\pi/T$ ($T = 1$ year) 
with $t_0 = 152.5$ days (from astronomical data).
Importantly the phase and period of the modulation are both
predicted!\footnote{
Deviations from $t_0 = 152.5$ days are possible if there is bulk halo
rotation. However, a large deviation from 
the expected value of $t_0 = 152.5$ days would be difficult to reconcile
with the inferred approximate
spherical distribution of the dark matter in the galactic halo.} 
This gives a strong systematic check on their results.
Such an annual modulation has been found in the 2-6 keVee recoil energy
region 
at the $8.9\sigma$ 
confidence level, with $T,\ t_0$ measured to be\cite{dama3}:
\begin{eqnarray}
T &=& 0.999 \pm 0.002 \ {\rm year} \nonumber \\
t_0 &=& 146 \pm 7 \ {\rm day.}
\end{eqnarray}
Clearly, both the period and phase are consistent with the theoretical
expectations of 
halo dark matter.  There are no known systematic effects which could
produce the 
modulation of the signal seen and thus it is reasonable to believe that
the 
DAMA experiments have detected dark matter.

Mirror dark matter explains the DAMA annual modulation signal via kinetic mixing
induced elastic (Rutherford) scattering of the dominant mirror metal component, $A'$, off
target nuclei. [The $H'$ and $He'$ components are too light to give a signal above
the DAMA energy threshold]. We leave it up to the experimental data to determine
$m_{A'}$, although our best `theoretical' guess would be $A' \sim O'$ given that $O$ is the
dominant metal in the ordinary matter sector.
The differential interaction rate is given
by:
\begin{eqnarray}
{dR \over dE_R} &=& 
 N_T n_{A'} \int {d\sigma \over dE_R} {f_{A'}({\bf{v}},{\bf{v}}_E) \over
k} |{\bf{v}}|
d^3v \nonumber \\
&=& N_T n_{A'}
{\lambda \over E_R^2 } \int^{\infty}_{|{\bf{v}}| > v_{min}
(E_R)} {f_{A'}({\bf{v}},{\bf{v}}_E) \over k|{\bf{v}}|} d^3 v 
\label{55}
\end{eqnarray}
where $N_T$ is the number of target atoms per kg of detector (we must sum over $Na$ and $I$
interactions separately),  
$k = (\pi v_0^2 [A'])^{3/2}$ is the Maxwellian distribution
normalization factor and
$n_{A'} = \rho_{dm} \xi_{A'}/m_{A'}$ is the number density of the mirror
nuclei $A'$ at the Earth's
location (we take $\rho_{dm} = 0.3 \ GeV/cm^3$).
Here ${\bf{v}}$ is the velocity of the halo particles relative to the
Earth and ${\bf{v}}_E$ is the
velocity of the Earth relative to the galactic halo.
Note that the lower velocity limit,
$v_{min} (E_R)$, 
is given by the kinematic relation:
\begin{eqnarray}
v_{min} &=& \sqrt{ {(m_A + m_{A'})^2 E_R \over 2 m_A m^2_{A'} }}
\ .
\label{v}
\end{eqnarray}

The velocity integral in Eq.(\ref{55}),
\begin{eqnarray}
I \equiv \int^{\infty}_{|{\bf{v}}| > v_{min}
(E_R)} {f_{A'}(v) \over k|{\bf{v}}|} d^3 v 
\label{55x}
\end{eqnarray}
can easily be evaluated in terms of error functions assuming a
Maxwellian distribution:
$f_{A'} ({\bf{v}},{\bf{v}}_E)/k = (\pi v_0^2[A'])^{-3/2} exp(-({\bf{v}}
+ {\bf{v}}_E)^2/v_0^2[A'])$.
In fact, 
\begin{eqnarray}
I = {1 \over 2yv_0[A']} [erf(x+y) - erf(x-y)]\ ,
\end{eqnarray}
where
\begin{eqnarray}
x \equiv {v_{min}(E_R) \over v_0[A']}, \ y \equiv {v_E \over v_0[A']} \
.
\end{eqnarray}

The differential interaction rate, Eq.(\ref{55}), can then be expanded
in a Taylor series yielding a time
independent part (which we subsequently denote as the `absolute' rate)
and time dependent modulated component:
\begin{eqnarray}
{dR \over dE_R} \simeq {dR^0 \over dE_R} + {dR^1 \over dE_R} \cos \omega
(t-t_0)\ ,
\end{eqnarray}
with
\begin{eqnarray}
{dR^0 \over dE_R} &=& {N_T n_{A'} \lambda I(E_R, y_0) \over E_R^2}
\nonumber \\
{dR^1 \over dE_R} &=& {N_T n_{A'} \lambda \Delta y \over E_R^2} \left(
{\partial I \over \partial y}\right)_{y=y_0}\ .
\end{eqnarray}
Here $y_0 = v_{\odot}/v_0[A'], \ \Delta y = \Delta v_E/v_0[A']$  and
\begin{eqnarray}
\left( {\partial I \over \partial y}\right)_{y=y_0} 
= - {I(E_R,y_0) \over y_0} + { 1 \over \sqrt{\pi} y_0 v_0 [A']} \left[
e^{-(x-y_0)^2} + e^{-(x+y_0)^2} \right]\ .
\end{eqnarray}

To compare with the
measured rates we must take into account the quenching factor and
detector resolution.
We include detector resolution effects by convolving the rates with a
Gaussian:
\begin{eqnarray}
{dR^{0,1} \over dE_R^m} =  {1 \over \sqrt{2\pi}\sigma_{res} } 
\int {dR^{0,1} \over dE_R} e^{-(E_R - E_R^m)^2/2\sigma^2_{res}} dE_R\ , 
\label{fri8}
\end{eqnarray}
where $E_R^m$ is the measured energy. The resolution is
given by\cite{damares} 
\begin{eqnarray}
{\sigma_{res} \over E_R} = {\alpha \over \sqrt{E_R (keVee)}} \ + \ \beta
\label{26}
\end{eqnarray}
where $\alpha = 0.448 \pm 0.035, \ \beta = (9.1 \pm 5.1) \times
10^{-3}$. 
The unit of energy is the electron equivalent energy, keVee.
For nuclear recoils, in the absence of any channeling,
keVee = keV/$q_A$, where $q_A$ is the quenching factor. For DAMA, $q_{Na}
\approx 0.3$ while
$q_I \approx 0.09$. Channeled events where target atoms travel down
crystal axis and planes have
$q_A \simeq 1$.

The issue of channeling has recently been re-examined in ref.\cite{gelmini}.
It was found that the channeling fraction is likely to be very small ($< 1\%$) in
the energy range of interest in contrast with the earlier study\cite{damachan} performed by the DAMA
collaboration. It is argued that the DAMA analysis did not take into account
that the scattered atoms originate from lattice sites and hence cannot be easily
channeled. In light of these developments, we expect that the 
channeling fraction is indeed small - probably negligible.
Throughout this paper, therefore,
we shall generally assume that no channeling occurs, with the exception of figures 2 and 4
where we compute the allowed region of parameter space and  
include the channeling region as a comparison.

The measured annual modulation amplitudes for the 1.17 ton-year cumulative
exposure, $S_i^m \pm \sigma_i$, are binned into
$\Delta E = 0.5$ keVee energy bins 
and can be obtained from figure 6 of ref.\cite{dama3}. This is to be
compared with the
computed annual modulation amplitude for mirror dark matter 
obtained by averaging the differential rate over the binned energy
taking into account the resolution and quenching factors:
\footnote{The DAMA data are efficiency corrected so there is no need to
include the detection efficiency in Eq.(\ref{sm}).}
\begin{eqnarray}
{\overline{dR}^1_i \over dE_R^m} = {1 \over \Delta E}\int^{E_i+\Delta
E}_{E_i} {dR^1 \over dE_R^m} dE_R^m\ .
\label{sm}
\end{eqnarray}
It is convenient to define a $\chi^2$ quantity:
\begin{eqnarray}
\chi^2 (\epsilon\sqrt{\xi_{A'}}, m_{A'}) = \sum \left({\overline{dR}^1_i
\over dE_R^m}
- S_i^m \right)^2/\sigma^2_i\ .
\end{eqnarray}
We consider the energy range 2-8 keVee, separated into 12
bins of width $0.5$ keVee, which
encompasses the 2-6 keVee DAMA signal region.
Varying the parameters $m_{A'}, \epsilon\sqrt{\xi_{A'}}$ around the best
fit, 
we can obtain the DAMA allowed region\footnote{
For the purposes of the fit, we analytically continue the mass number,
$A'$, to non-integer values,
with $Z' = A'/2$. Since the realistic case will involve a spectrum of
elements, the effective
mass can be non-integer}. 
There are a number of systematic uncertainties which can be included in
the analysis and we have examined the following: 
a) considering $v_0[A']$ within its expected limits
given by Eq.(\ref{limits2}), b) varying the quenching factors for Iodine
and sodium
by $\pm 20\%$, i.e. taking $q_{Na} = 0.30 \pm 0.06$ and $q_I = 0.09 \pm
0.02$, 
c) varying the detector resolution over its $2\sigma$ uncertainty 
and d) varying $v_{rot} = 254 \pm 32$ km/s, that is around $\pm 2\sigma$
from its estimated value. 
The variation of quenching factor and resolution were taken into account by
minimizing $\chi^2 (q_I, q_{Na}, \sigma_{res}, m_{A'},
\epsilon\sqrt{\xi_{A'}})$ 
over $20\%$ variation of $q_I$ and $q_{Na}$, and over the 
$2\sigma$ uncertainty in $\sigma_{res}$. This defines 
$\stackrel{-}{\chi}^2 (m_{A'}, \epsilon\sqrt{\xi_{A'}})$. 
The best fit for DAMA has $\stackrel{-}{\chi}^2_{min} \simeq 8.5$ for 10 degrees of freedom.
Some examples near the best fit, assuming no channeling occurs, 
are shown in figure 1.

\vskip 0.3cm
\centerline{\epsfig{file=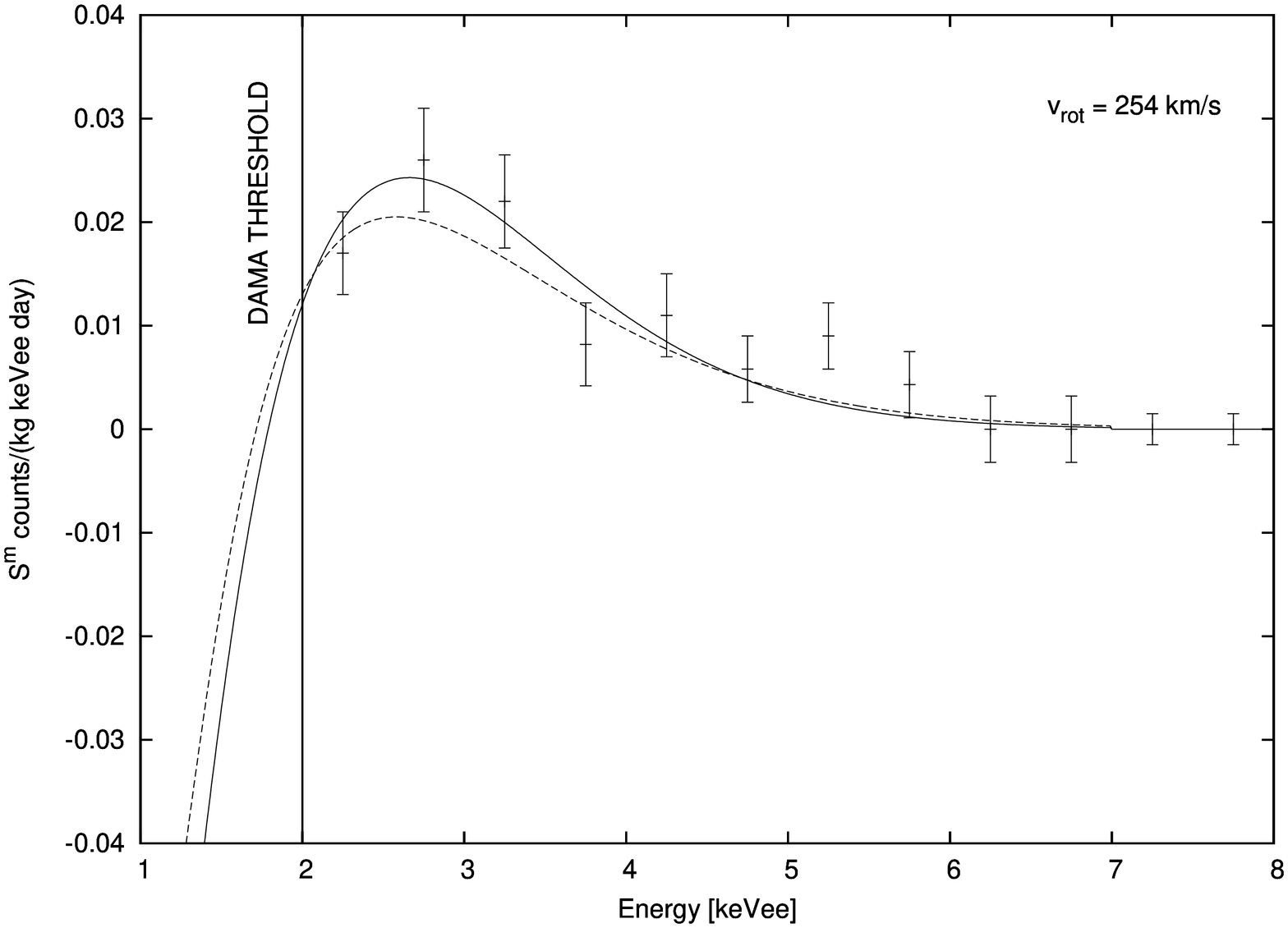,angle=270,width=12.8cm}}
\vskip 0.5cm
\noindent
Figure 1: DAMA annual modulation amplitude  
versus measured recoil Energy for the parameters:
$m_{A'}/m_p = 20, \ \epsilon \sqrt{\xi_{A'}} = 7.4\times 10^{-10}$ and
$v_{rot} = 254$ km/s. The solid (dashed) line corresponds to the lower (upper)
$v_0[A']$ limit given in Eq.(\ref{limits2}). 
Negligible channeling has been assumed.
\vskip 1.0cm

Favoured regions in the $m_{A'}, \epsilon\sqrt{\xi_{A'}}$ plane can be obtained by 
evaluating the contours 
with $\stackrel{-}{\chi}^2 = \stackrel{-}{\chi}^2_{min} + 9$ (roughly 99\% C.L.).  
In figure 2 we plot the allowed regions for DAMA 
%together with allowed region for Cogent 
assuming $v_{rot} = 222$ km/s [fig 2a], $v_{rot} = 254$ km/s [fig 2b] and
$v_{rot} = 286$ km/s [fig 2c] .
%Also shown are the  corresponding $95\%$ C.L. exclusion limits from
%Xenon100, CDMS/Si and CDMS/Ge.
Also shown in the figures are the DAMA allowed region assuming that
channeling occurs with fractions
as originally estimated by the DAMA collaboration\cite{damachan}.
It has been emphasised recently\cite{cogent2} that the systematic uncertainty 
in $q_{Na}$ might be as
large as $q_{Na} = 0.30 \pm 0.13$ given the lack of measurements of the quenching factor
in the low energy region. If this is the case, then we find that the favoured regions extend out to
somewhat ($\approx 10-15\%$) lower $m_{A'}$ values than that given in figure 2.
%\vskip 0.1cm
%\centerline{\epsfig{file=fig3n.eps,angle=270,width=12.4cm}}
%\vskip 0.3cm
%\noindent
\vskip 0.1cm
\centerline{\epsfig{file=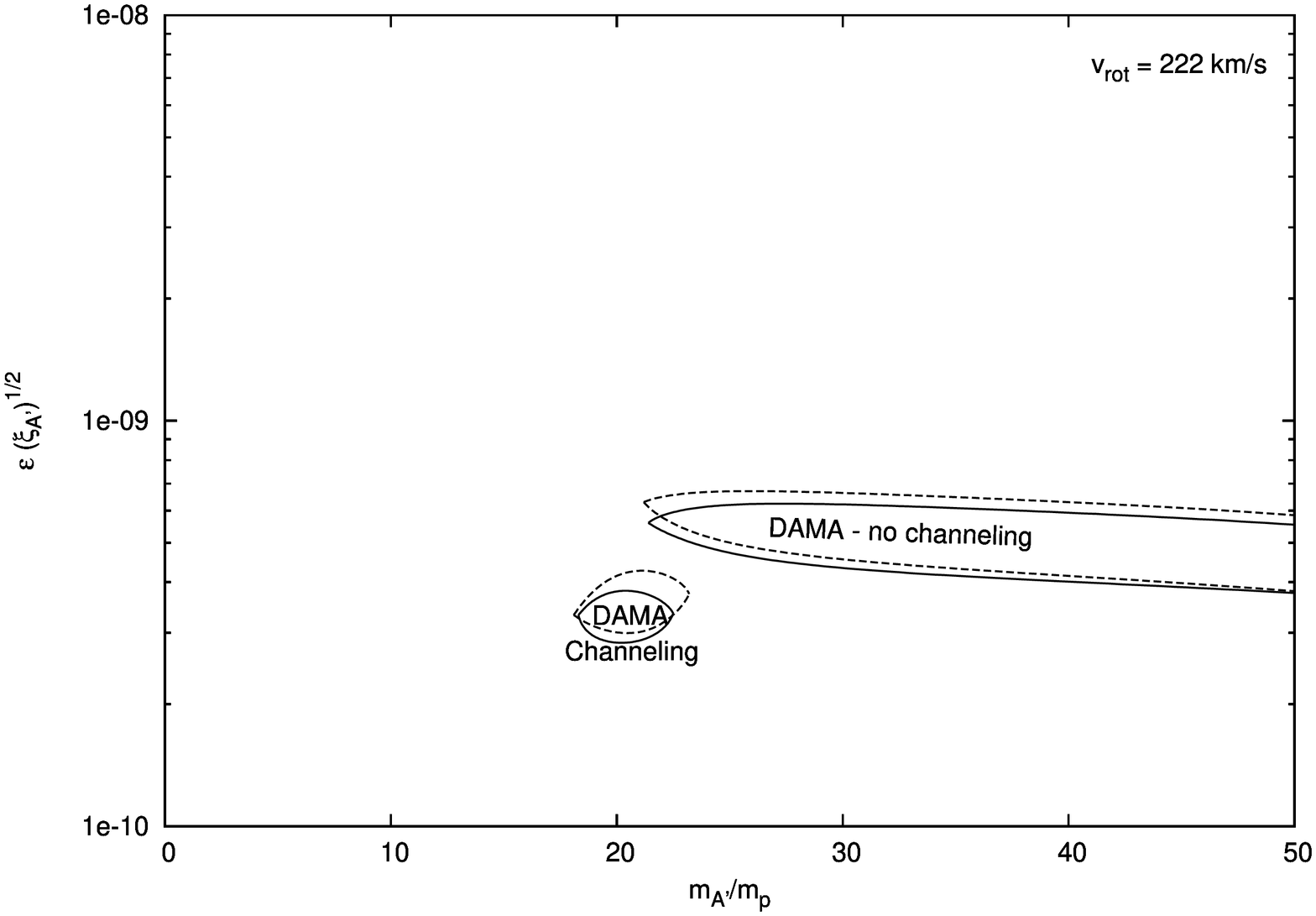,angle=270,width=12.4cm}}
\vskip 0.3cm
\noindent
Figure 2a: DAMA [99\% C.L.] allowed region in the $m_{A'}, 
\epsilon \sqrt{\xi_{A'}}$ plane, assuming negligible channeling fraction,
for $v_{rot} = 222$ km/s. 
Solid (dashed) line corresponds to the lower (upper) $v_0[A']$ limit
given in Eq.(\ref{limits2}).
Also shown for comparison is the DAMA allowed regions if channeling occurs with
fractions originally estimated
by the DAMA collaboration.
\vskip 0.1cm
\centerline{\epsfig{file=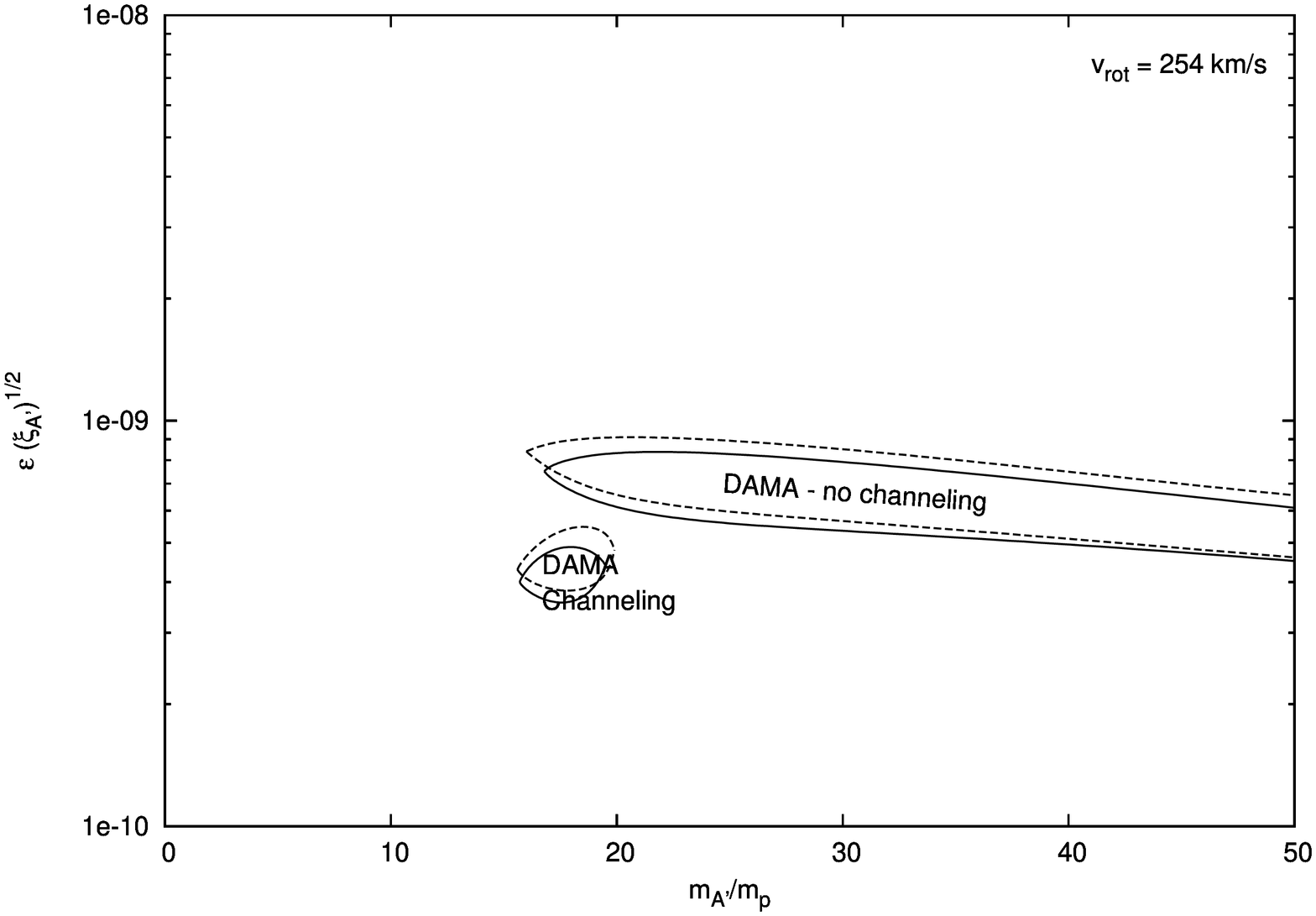,angle=270,width=12.4cm}}
\vskip 0.4cm
\noindent
Figure 2b: Same as figure 2a, except with $v_{rot} = 254$ km/s.
\vskip 0.4cm
\centerline{\epsfig{file=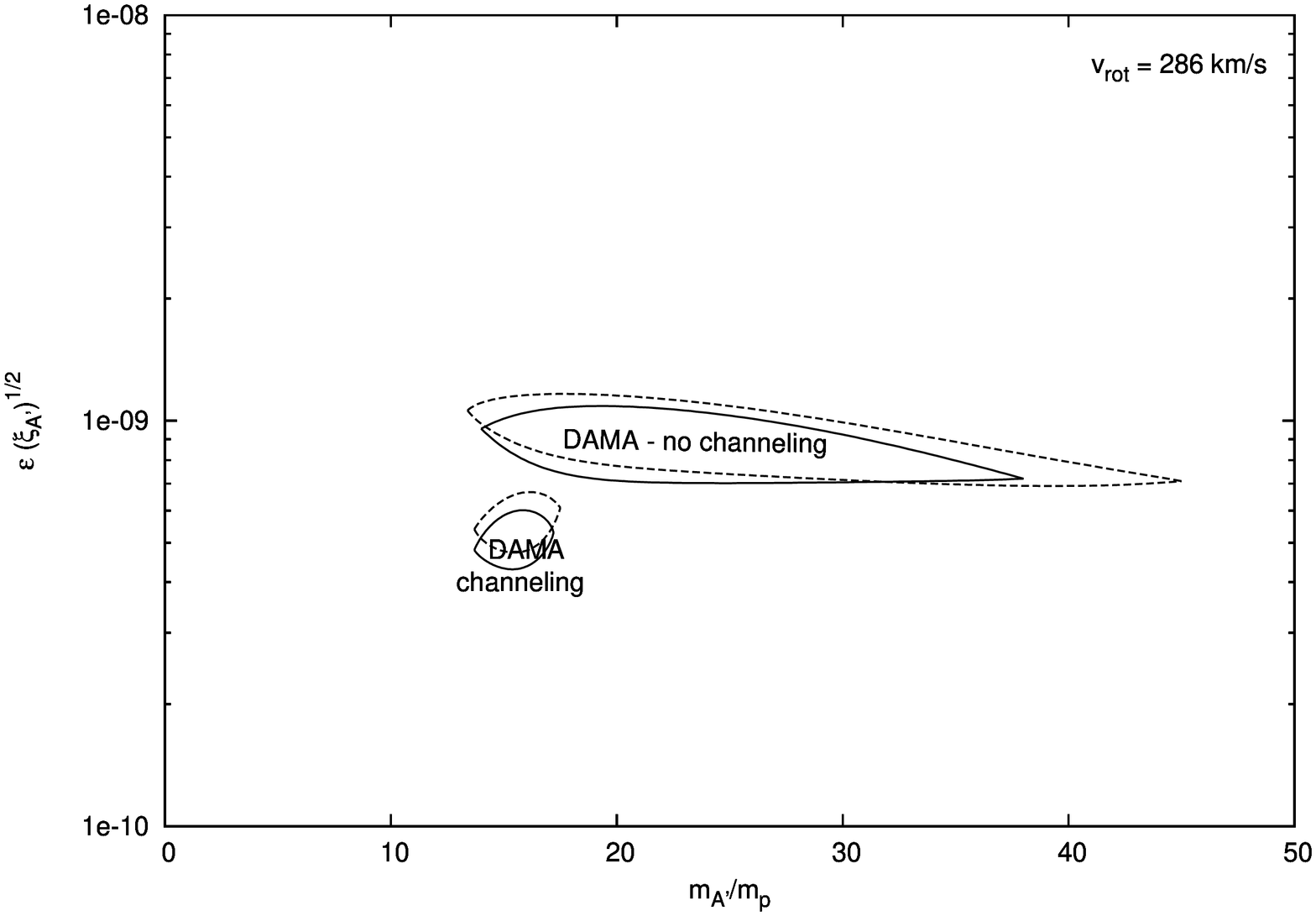,angle=270,width=12.4cm}}
\vskip 0.4cm
\noindent
Figure 2c: Same as figure 2a, except with $v_{rot} = 286$ km/s.

\vskip 1.0cm

%\vskip 0.1cm
%\centerline{\epsfig{file=fig3n.eps,angle=270,width=12.4cm}}
%\vskip 0.3cm
%\noindent
%Figure 3: DAMA [99\% C.L.] allowed region in the $m_{A'}, 
%\epsilon \sqrt{\xi_{A'}}$ plane, assuming negligible channeling fraction,
%for $v_{rot} = 254$ km/s. The thick line assumes $q_{Na} = 0.30 \pm 0.06$, 
%while the thin line assumes
%$q_{Na} = 0.30 \pm 0.13$. [A $v_0[A']$ value given by the lower limit
%in Eq.(\ref{limits2}) is assumed.]
%\vskip 1cm

The main features of the annual modulation spectrum predicted by $A'$
dark matter can be easily understood. As discussed 
earlier\cite{mm2}, at low $E_R$, where $x(E_R) \ll y$,
$dR^1/dE_R$ is negative. As $E_R$ increases, $dR^1/dE_R$ changes sign
and reaches a maximum at the value of $E_R$ where 
$x(E_R) \approx y$, or equivalently, the value
of $E_R$ where $v_{min}(E_R) = v_{rot}$. At high $E_R$ ($x \gg y$),
$dR^1/dE_R \to 0$. From Eq.(\ref{v}) this means that the position 
of the peak, $E_R^{peak}$, is given by:
\begin{eqnarray}
E_R^{peak} \approx {2 m_{A} m_{A'}^2 \over (m_{A} + m_{A'})^2} v_{rot}^2 \ .
\end{eqnarray}
For low $m_{A'} \stackrel{<}{\sim} 30 m_p$ the annual modulation signal
arises predominantly from $A'$ scattering with $Na$, 
while for $m_{A'} \stackrel{>}{\sim} 30 m_p$ scattering off
both $Na$ and $I$ contributes significantly to produce the
signal\footnote{
In the case where channeling is assumed with the fractions originally
estimated by the DAMA collaboration\cite{damachan} the annual 
modulation signal is dominated by interactions with $I$ only. 
See ref.\cite{mm2} for further discussion of this case.}.
Note that since $v_0 [A'] \ll v_{rot}$ the velocity distribution
is so narrow that the width of the peak is dominated by the detector
resolution. This explains the relative insensitivity of the fit 
to the particular $v_0 [A']$ value which is evident in figures 1,2.

To summarize,
we see that without channeling the DAMA
annual modulation signal
can be explained for a relatively wide range of $m_{A'}$ values, which
includes the region 
around $m_{A'}/m_p \sim 16$, expected if $A' = O'$ dominates the
mirror metal component. 
We now turn to the positive low energy excess observed by CoGeNT,
together with the implications
from the null experiments, such as XENON100 and CDMS/Si.

\section{The CoGeNT experiment}

The CoGeNT experiment operating in the Soudan Underground Laboratory has 
recently presented new results in their search for light dark matter
interactions\cite{cogent}.
With a low energy threshold of $0.4$ keVee and a Germanium target, they
have observed a low energy excess which
is not easily explainable in terms of known background sources. The
energy region probed by CoGeNT
overlaps with the energy region in which the DAMA collaboration have
observed their impressive annual modulation
signal and thus it is natural to interpret the CoGeNT low energy excess
in terms of $A'$ mirror dark matter
interactions. In ref.\cite{mmcogent} we showed that the CoGeNT excess is
compatible with mirror dark matter
expectations and thus provides a model dependent check of the DAMA
signal, which we now examine in
more detail. 

To compare with the measured event rate, we include detector resolution
effects and overall detection 
efficiency:
\begin{eqnarray}
{dR^0 \over dE_R^m} = \epsilon_f (E_R^m) {1 \over \sqrt{2\pi}\sigma } 
\int {dR^0 \over dE_R} e^{-(E_R - E_R^m)^2/2\sigma^2} dE_R 
\end{eqnarray}
where $E_R^m$ is the measured energy and $\sigma^2 = \sigma_n^2 +
(2.35)^2 E_R \eta F$
with $\sigma_n = 69.4$ eV, $\eta = 2.96$ eV and $F
=0.29$\cite{cogent,cog2}.
The detection efficiency, $\epsilon_f (E_R^m)$, was given in figure 3 of
ref.\cite{cogent},
which we approximate via
\begin{eqnarray}
\epsilon_f (E_R^m) \simeq {0.87 \over 1 + (0.4/E_R^m)^{6} }.
\end{eqnarray}
The energy is in keVee units (ionization energy). For nuclear recoils
in the absence of any channeling, $keVee = keV/q$, where $q \simeq 0.21$
is the 
relevant quenching factor in the near threshold region\cite{cogent2}.

We fit the CoGeNT data in the low recoil energy range assuming $A'$
dark matter and that the background is an energy independent constant,
together with two Gaussians
to account for the $^{65}Zn$ (1.1 keV) and $^{68}Ge$ (1.29 keV) 
L-shell electron capture lines.
Initially fixing $m_{A'}/m_p = 20$ and $v_{rot} = 254$ km/s, as an example, we
find a best fit
of $\chi^2_{min} \simeq 14.8$ for $21-4 = 17$ degrees of freedom, with
$\epsilon \sqrt{\xi_{A'}} = 6.2\times 10^{-10}$
(independently of whether we take the upper or lower limiting values of $v_0$).
This fit for CoGeNT is  shown in figure 3. 
\vskip 1cm
\centerline{\epsfig{file=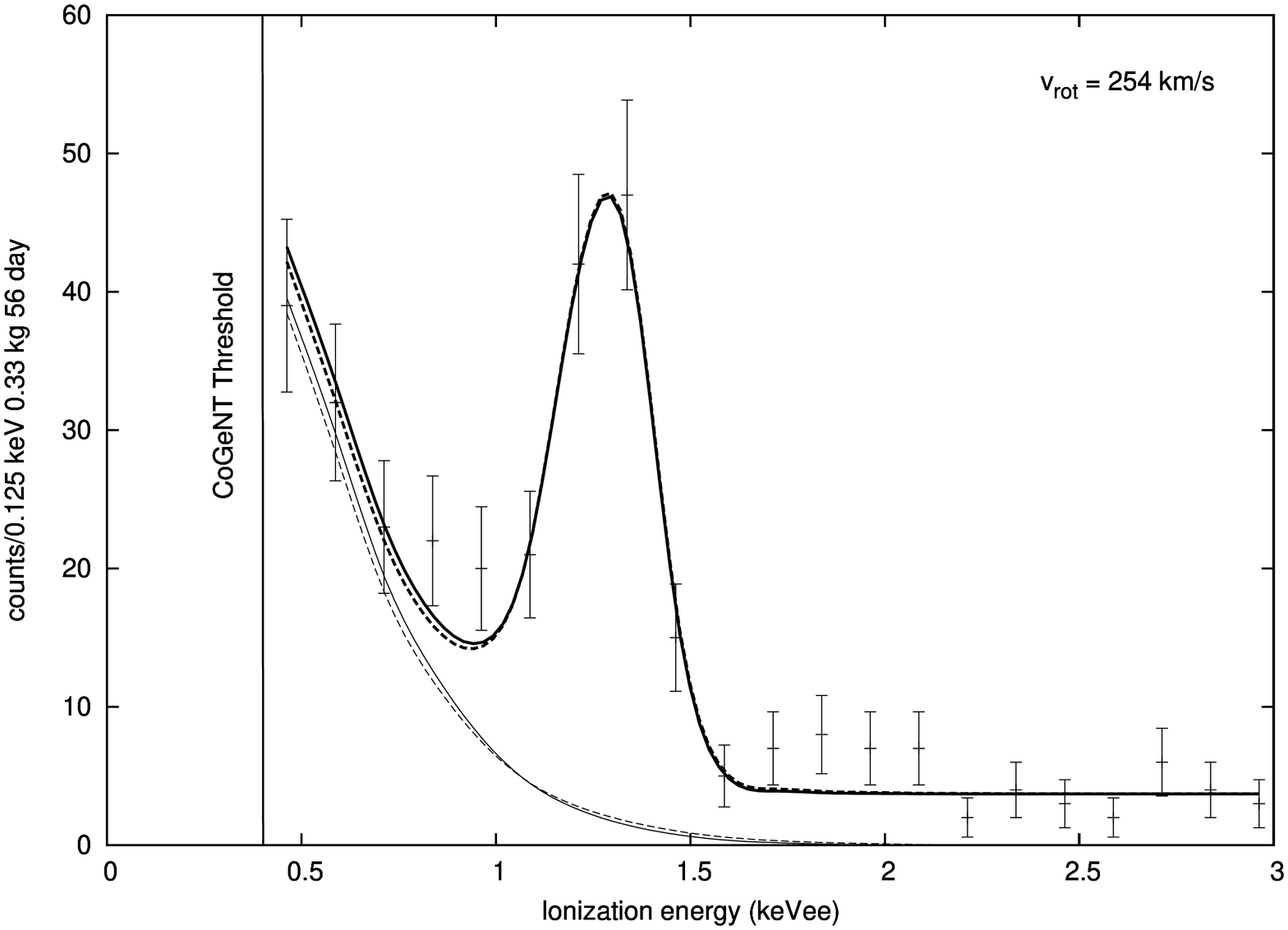,angle=270,width=12.7cm}}
\vskip 0.5cm
\noindent
Figure 3: Fit of the 
CoGeNT spectrum for $m_{A'}/m_p = 20, \ \epsilon \sqrt{\xi_{A'}} =
6.2\times 10^{-10}$ 
and $v_{rot} = 254$ km/s.
Thick solid (dashed) line is with lower (upper) $v_0[A']$ limit given in
Eq.(\ref{limits2}).
Also shown is the dark matter contribution to the signals (thin lines).
Negligible channeling has been assumed.

\vskip 1.2cm

The shape of the spectrum is nicely fit by $A'$ dark matter due
primarily to the $E_R$ dependent Rutherford cross section:
$d\sigma/dE_R \propto 1/E_R^2$. As further data is collected this
$E_R$ dependence will be more stringently constrained, which
will pose a more rigorous test of the mirror dark matter theory.

As in the DAMA case,
to account for some of the possible systematic uncertainties
we vary the quenching factor by $\pm 20\%$ [i.e. take $q_{Ge} = 0.21\pm
0.04$].
%and consider a range of $v_{rot}$ and halo dispersion 
%velocity [Eq.(\ref{limits2})]. 
We define a $\chi^2$ function,
\begin{eqnarray}
\chi^2 (q_{Ge},\epsilon\sqrt{\xi_{A'}}, m_{A'}) = \sum \left(
{\overline{dR}_i^0 \over dE_R^m} - data_i \right)^2/\sigma^2_i
\end{eqnarray}
where ${\overline{dR}_i^0 \over dE_R^m}$ is the differential rate
averaged over the 
binned energy. We take the experimental errors to be purely statistical,
so that $\sigma_i = \sqrt{data_i}$.
We minimize 
$\chi^2 (q_{Ge}, m_{A'}, \epsilon\sqrt{\xi_{A'}})$ 
over the $20\%$ variation in $q_{Ge}$, which defines 
$\stackrel{-}{\chi}^2 (m_{A'}, \epsilon\sqrt{\xi_{A'}})$.
Of course, we also minimize $\chi^2$ with respect
to the parameters of the background model describing the amplitudes of
the $^{65}Zn$ (1.1 keV) and $^{68}Ge$ (1.29 keV) 
L-shell electron capture lines and adjusting also the constant
background component.  
Note that no background exponential is assumed (or needed) to fit the
data. 
The CoGeNT allowed region in the $m_{A'}, \epsilon\sqrt{\xi_{A'}}$ plane
is then defined via contours 
$\stackrel{-}{\chi}^2 = \stackrel{-}{\chi}^2_{min} + 9$ (roughly 99\%
C.L.). 

In figure 4 we plot the allowed regions 
for CoGeNT together with the DAMA allowed region 
for $v_{rot} = 222$ km/s [fig 4a], $v_{rot} = 254$ km/s [fig 4b] and
$v_{rot} = 286$ km/s [fig 4c] . We have allowed for  a $20\%$ systematic uncertainty in both
CoGeNT and DAMA quenching factors.
In these figures we have assumed  $\xi_{A'} \ll 1$ so that
$v_0[A']$ is given by the lower limit
given in Eq.(\ref{limits2}). The case where $\xi_{A'} \simeq 1$ features an almost
indistinguishable allowed region (as already illustrated in figure 2 for the case of DAMA),
so is not given. 
Also shown are the  corresponding $95\%$ C.L. exclusion limits from
CDMS/Si\cite{cdmssi}, CDMS/Ge\cite{cdmsge}
and XENON100\cite{xenon100} experiments.

\vskip 1.2cm
\centerline{\epsfig{file=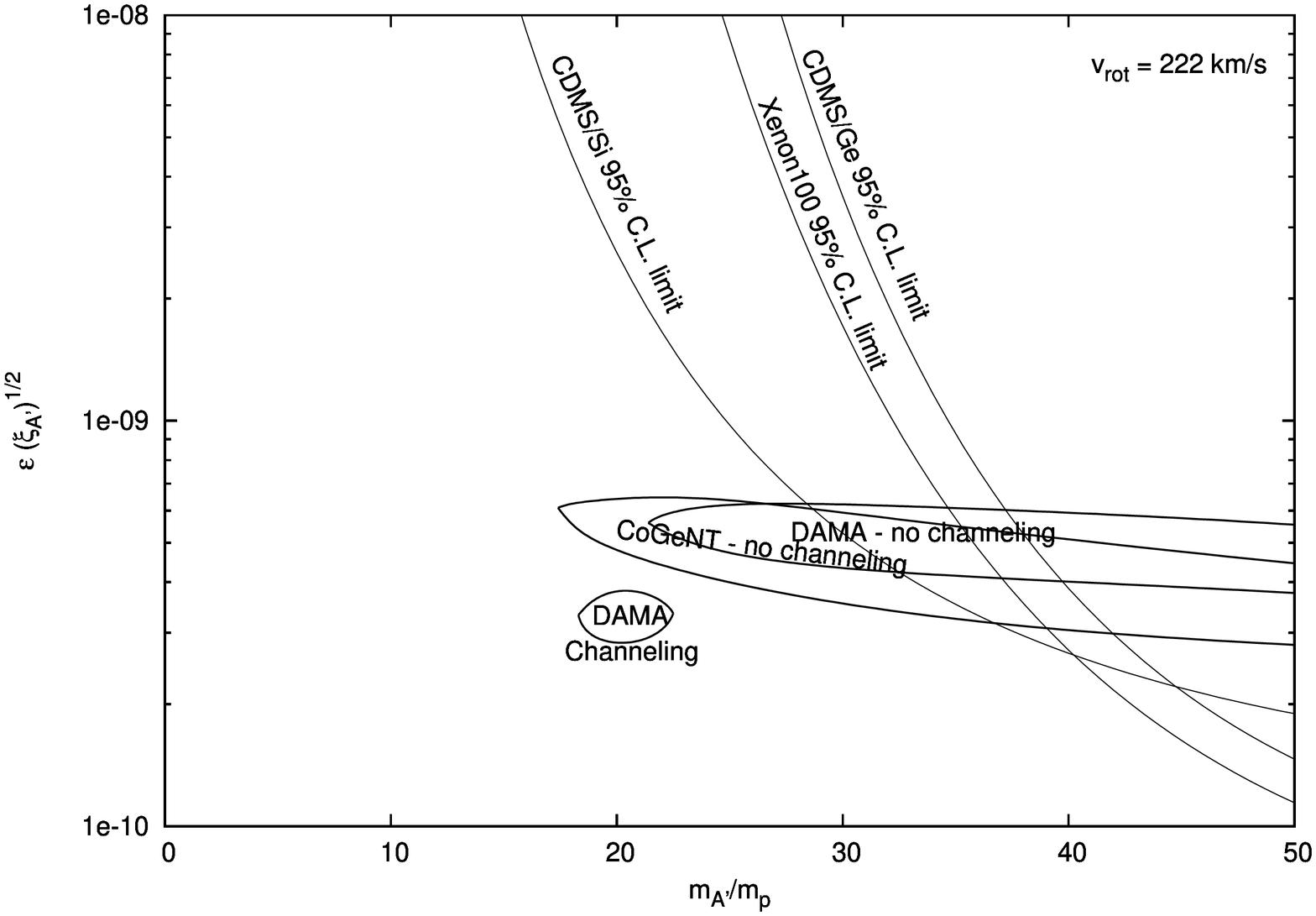,angle=270,width=12.8cm}}
\vskip 0.3cm
\noindent
Figure 4a: DAMA and CoGeNT [99\% C.L.] allowed regions in the 
$m_{A'}/m_p, \ \epsilon \sqrt{\xi_{A'}}$ plane, 
assuming negligible
channeling fraction, for $v_{rot} = 222$ km/s. 
%The solid line assumes $q_{Na} = 0.30 \pm 0.06$ while the dashed line assumes $q_{Na} = 0.30 \pm 0.13$.
Also shown are a) DAMA allowed regions if channeling occurs with
fractions originally estimated
by the DAMA collaboration and b) the exclusion limits from CDMS/Si,
CDMS/Ge and XENON100 experiments. [The region excluded is to the right of the exclusion limits]. 

\vskip 0.2cm

\centerline{\epsfig{file=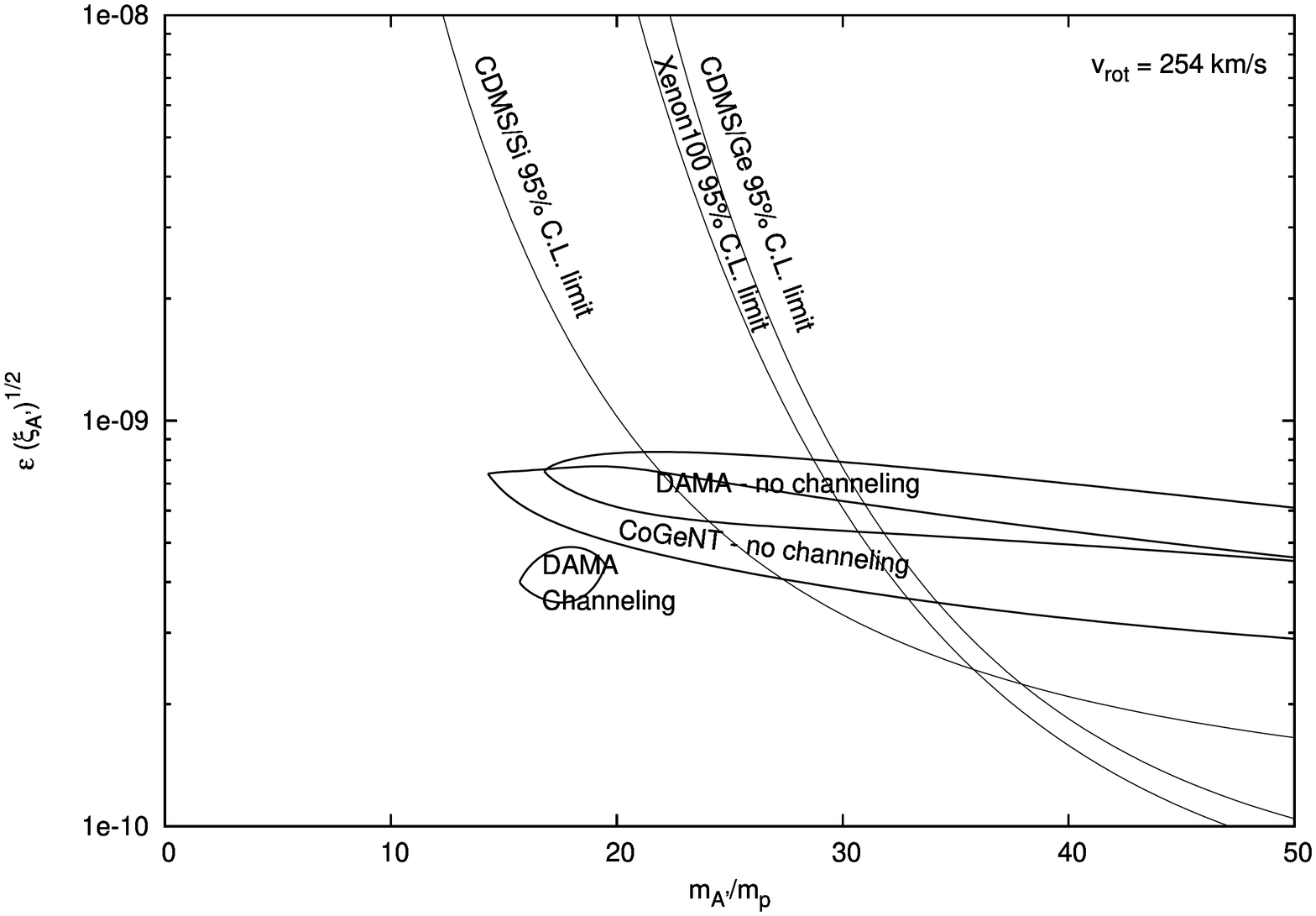,angle=270,width=12.8cm}}
\vskip 0.4cm
\noindent
Figure 4b:  Same as figure 4a, except with $v_{rot} = 254$ km/s.
\vskip 0.8cm
\centerline{\epsfig{file=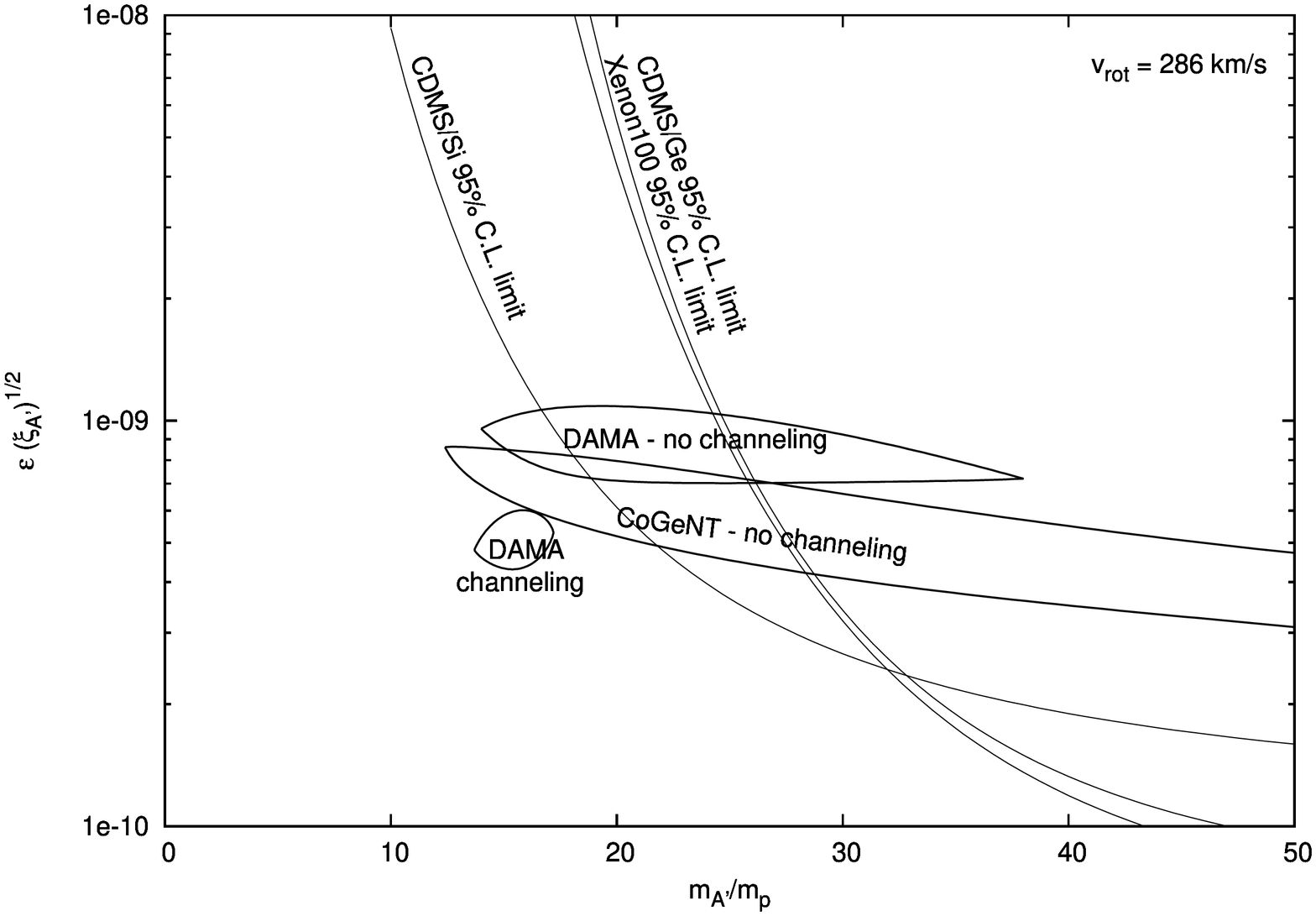,angle=270,width=12.8cm}}
\vskip 0.4cm
\noindent
Figure 4c: Same as figure 4a, except with $v_{rot} = 286$ km/s.
\vskip 1.2cm

In computing the exclusion limits we have allowed for a $20\%$ 
systematic uncertainty in energy threshold.
That is, the threshold for CDMS/Si\cite{cdmssi} was taken to be $8.4\
keV_{nr}$ rather 
than the quoted $7.0\ keV_{nr}$, the threshold for CDMS/Ge\cite{cdmsge}
was taken to be
$12 \ keV_{nr}$ rather than the quoted $10\ keV_{nr}$ and
%the threshold for XENON10 was taken to be xxx rather than the quoted
%xxxx
the threshold for XENON100\cite{xenon100} was taken to be $10.4\
keV_{nr}$ 
rather than the quoted $8.7\ keV_{nr}$\footnote{Exclusion limits from
the XENON10 experiment are 
only marginally better than the XENON100 exclusion limit (but not better than the CDMS/Si limit). 
For a recent discussion about the calibration and other uncertainties in the XENON low
energy region, see ref.\cite{collar}.}.
For the CDMS/Si experiment, we used the quoted\cite{cdmssi} raw exposure
of $33.9$ kg-days for 
$E_R < 10\ keV_{nr}$, $38.8$ kg-days for $10 < E_R (keV_{nr}) < 15$ and
$53.5$ kg-days 
for $E_R > 15\ keV_{nr}$, and assumed a detection efficiency of $20\%$
in the energy region 
near threshold. 
%For the XENON10 experiment, we used the quoted\cite{xenon10} raw
%exposure 
%of  xxxx kg-days and detection efficiency taken from red curve in slide
%15 of Ref.\cite{sorenson} 
%[which we analytically approximated via $epsilon_f \approx 1/(1 +
%(4.8/E_R)^4)$].
For the CDMS/Ge experiment we assumed the total raw exposure of $1010$
kg-days and 
a detection efficiency of $\epsilon_f = 0.18 + 0.007E_R$ in the low
energy region of interest\cite{cdmsge}.
For the XENON100 experiment we used the quoted\cite{xenon100} raw
exposure of $447$
kg-days together with a detection efficiency of $0.4$ in the low energy
region.

From figure 4 it is evident that the CDMS/Si experiment is the most sensitive
of the null experiments to $A'$ dark matter, which
is due to its light target element and relatively low threshold. The
CDMS/Si experiment 
constrains the allowed region to $m_{A'}/m_p \stackrel{<}{\sim} 30$.
Observe also that the CDMS/Si experiment seems to exclude the two events
seen in the  
CDMS/Ge experiment\cite{cdmsge} as being due 
to the same component which can explain the DAMA and CoGeNT data.
It is possible, though, to interpret the two events seen by CDMS/Ge as a
hint of a heavier $\sim Fe'$ component.
This interpretation requires\cite{mmcdms} $\xi_{Fe'}/\xi_{A'} \sim
10^{-2}$ and is not excluded by CDMS/Si or any other experiment.
Such a small $Fe'$ component does not
significantly affect the fit of the DAMA or CoGeNT experiments.
We will examine the effects of the $Fe'$ component for the higher threshold experiments
in more detail in section 7.

We have also performed a global analysis of the DAMA and CoGeNT signals.
Fixing $v_{rot} = 254$ km/s and evaluating $\chi^2$ for the combined DAMA+CoGeNT data, we have found 
$\chi^2_{min} = 24.4$ for 28 d.o.f.
at $m_{A'}/m_p = 24, \ \epsilon \sqrt{\xi_{A'}} = 6.4\times 10^{-10}$.
Excellent fits to the combined DAMA and CoGeNT data are also obtained
for $v_{rot} = 222$ km/s and $v_{rot} = 286$ km/s.
In figure 5 we plot the favoured  region 
of parameter space for the combined fit of the DAMA and CoGeNT data for three representative
values of $v_{rot}$.

\vskip 0.2cm
\centerline{\epsfig{file=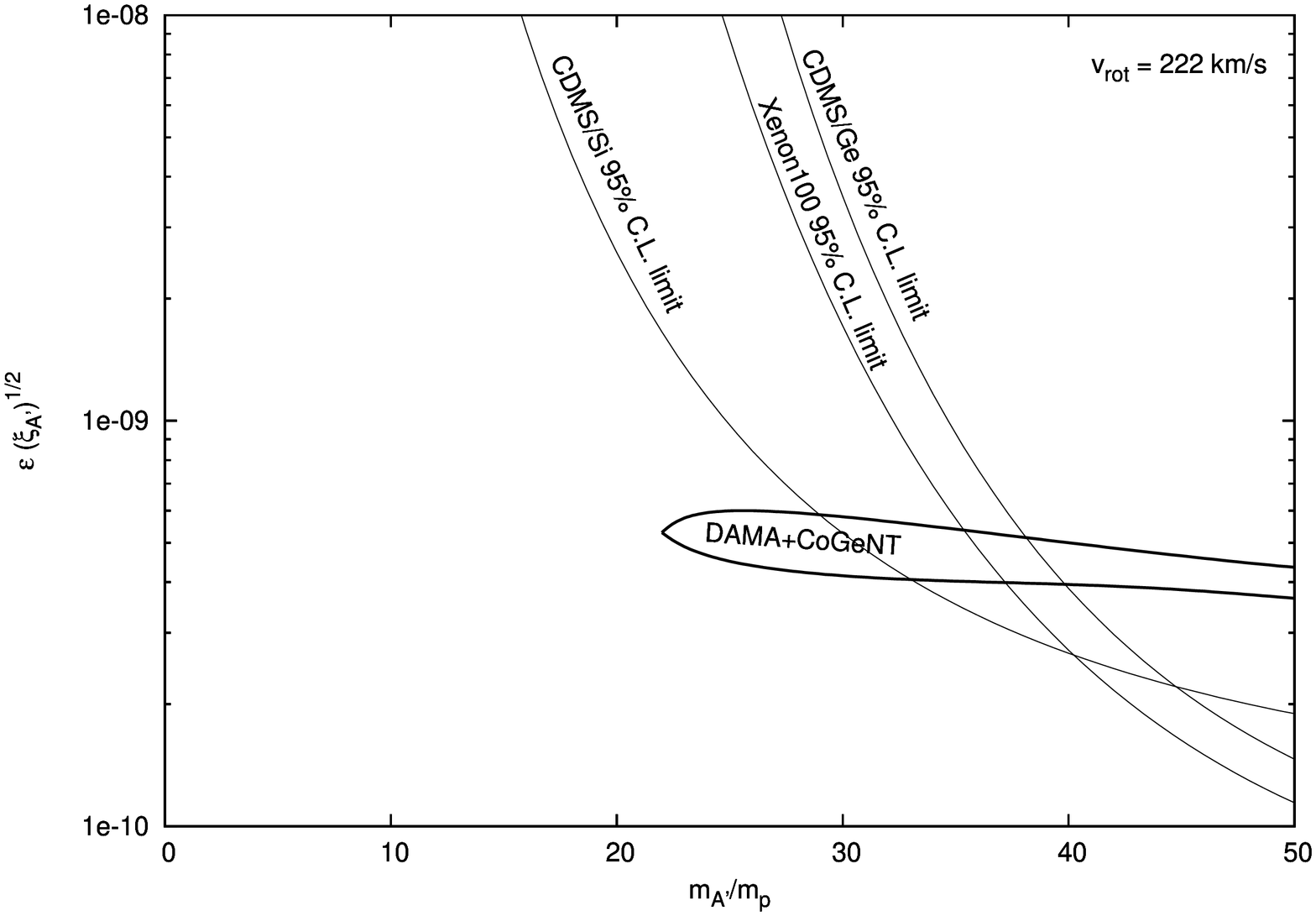,angle=270,width=12.6cm}}
\vskip 0.3cm
\noindent
Figure 5a: DAMA and CoGeNT [99\% C.L.] global allowed region in the 
$ m_{A'}/m_p, \ \epsilon \sqrt{\xi_{A'}}$ plane, 
assuming negligible
channeling fraction, for $v_{rot} = 222$ km/s. 
%The thick solid line assumes $q_{Na} = 0.30 \pm 0.06$ while the thin solid line
%assumes $q_{Na} = 0.30 \pm 0.13$. 
%Solid (dashed) line corresponds to the lower (upper) $v_0[A']$ limit
%given in Eq.(\ref{limits2}).
Also shown are the exclusion limits from CDMS/Si,
CDMS/Ge and XENON100 experiments. [The region excluded is to the right of the exclusion limits]. 
\vskip 0.7cm
\centerline{\epsfig{file=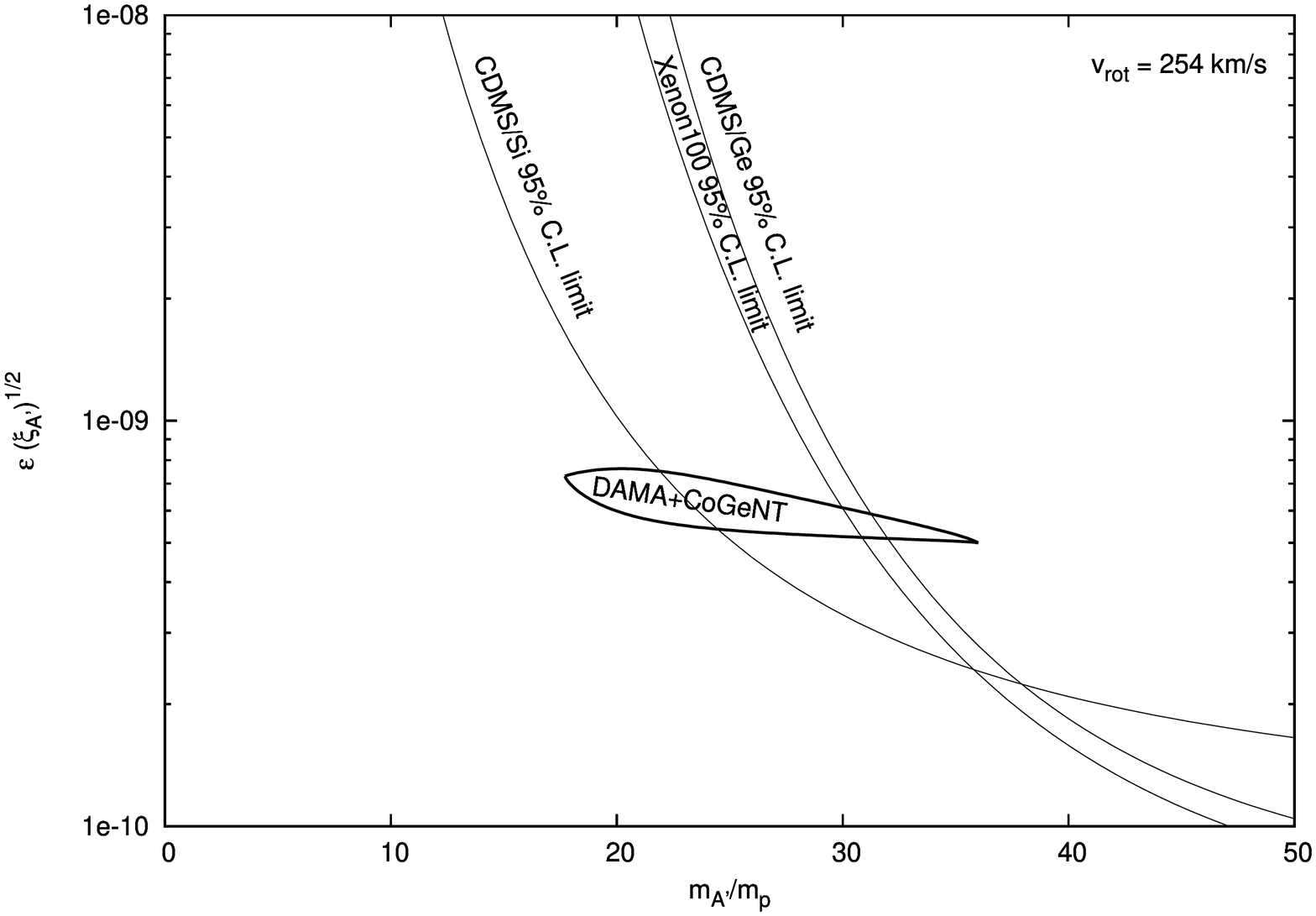,angle=270,width=12.6cm}}
\vskip 0.3cm
\noindent
Figure 5b: Same as figure 5a except that $v_{rot} = 254$ km/s.
\vskip 1cm
\centerline{\epsfig{file=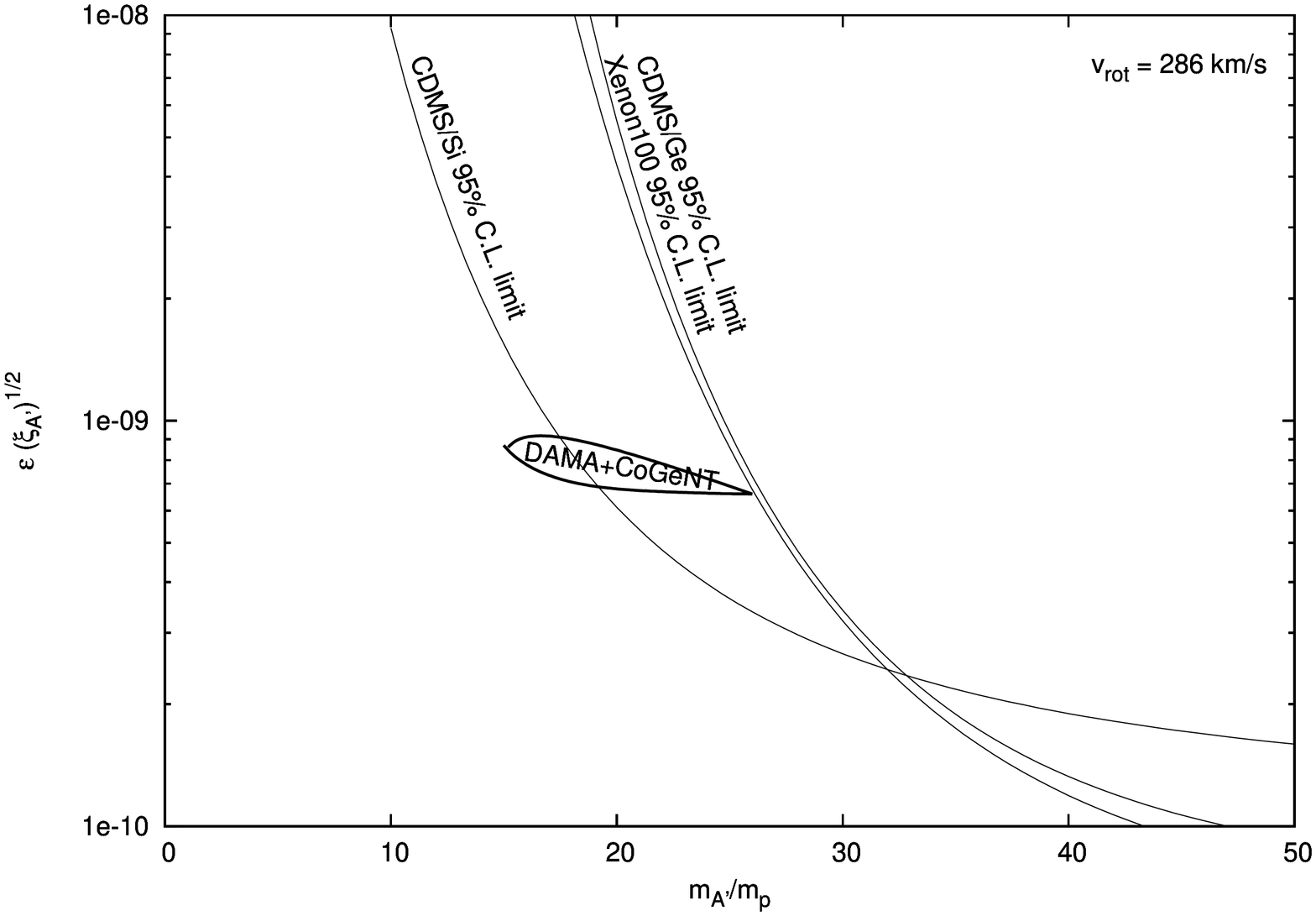,angle=270,width=12.8cm}}
\vskip 0.3cm
\noindent
Figure 5c: Same as figure 5a except that $v_{rot} = 286$ km/s.
\vskip 1cm  

The CoGeNT experiment has been running continuously since December 2009, 
and might collect
enough data to search for the annual modulation signal. 
In figure 6 we show results for the annual modulation
amplitude predicted for the CoGeNT experiment. Interestingly we see that there is a change 
of sign for the
annual modulation amplitude at low energies $E_R \approx 0.5-0.8$ keVee
(depending on the parameters). 
This means that at the lowest energies CoGeNT
should see {\it more} events during the (northern) winter/fall than the (northern)
summer/spring. This might
provide a useful means of experimentally distinguishing this 
dark matter theory from other possible explanations. 
\vskip 0.2cm
\centerline{\epsfig{file=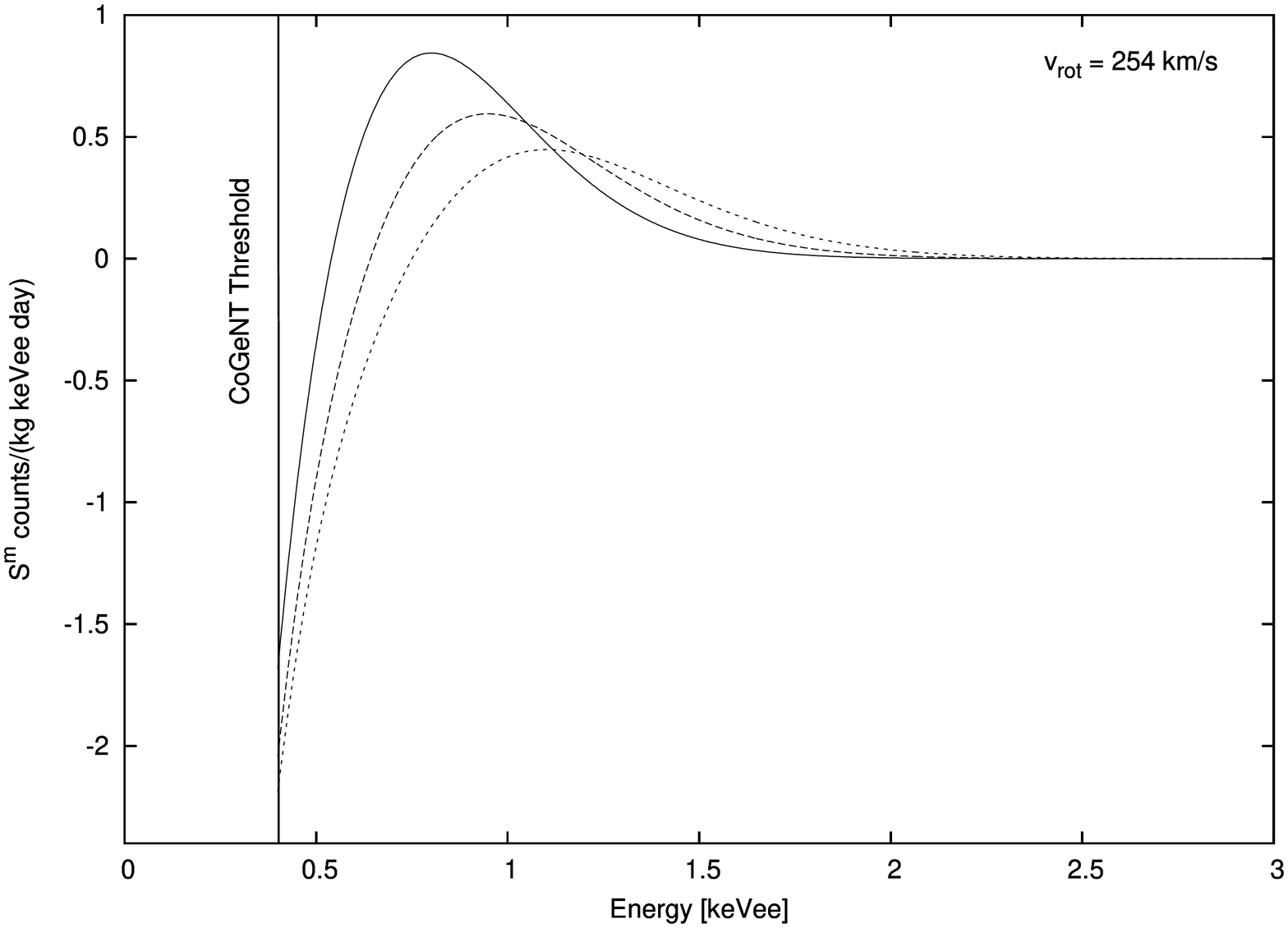,angle=270,width=12.0cm}}
\vskip 0.3cm
\noindent
Figure 6: CoGeNT annual modulation amplitude versus recoil energy.
For some representative parameters: $\epsilon \sqrt{\xi_{A'}} = 7.0\times 10^{-10}, \ 
m_{A'}/m_p = 18$ (solid line), 
$\epsilon \sqrt{\xi_{A'}} = 6.4\times 10^{-10}, \ 
m_{A'}/m_p = 20$ (dashed line), 
$\epsilon \sqrt{\xi_{A'}} = 6.0\times 10^{-10}, \ 
m_{A'}/m_p = 22$ (dotted line). All for $v_{rot} = 254$ km/s. The figure assumes
$100\%$ detection efficiency.  
 
\section{Electron scattering and the DAMA absolute rate}

In addition to nuclear recoils, electron recoils from mirror electron
scattering off bound
atomic electrons in the target volume can also occur.
The $e'$ component of the halo is expected to be distributed with
Maxwellian velocity distribution,
and since $m_{e'} \ll \bar m$, we expect $v_0[e'] \gg v_{rot}$.
In fact, from Eqs.(\ref{dis},\ref{mary}) we can estimate that $v_0[e']$ is
in the range:
\begin{eqnarray}
\left[ {m_p \over m_e}\right] {1 \over 2 - \frac{5}{4} \xi_{He'}}
\stackrel{<}{\sim}
{v^2_0 [e'] \over v_{rot}^2 } \stackrel{<}{\sim} 
\left[ {m_p \over m_e}\right] {A' \over 1 + A'/2} \ .
\label{fed}
\end{eqnarray}
Since $v_0[e'] \gg v_{rot}$ we can, to a good approximation, neglect
$v_{rot}$ when dealing
with $e'-e$ scattering. This means that electron recoils give a
negligible contribution to
the annual modulation signal, however they can contribute significantly
to the absolute
rate at energies below 2 keVee. Interestingly such a rise at low
energies was 
observed in the CDMS electron scattering data\cite{cdmselectron} and is
compatible with 
$\epsilon \sim 10^{-9}$\cite{mmelectron}. Recall that the DAMA
experiment doesn't
discriminate against electron recoils, so 
the DAMA experiment is also sensitive to electron recoils through their
contribution to the absolute
rate as we now discuss.

The DAMA experiment constrains the absolute event rate to be less than
around
1 cpd/kg/keVee in the region near threshold. We shall here examine the
implications of this
constraint for mirror dark matter. 
As discussed above, we expect both nuclear recoils and electron recoils
to contribute
to the absolute rate from dark matter interactions. 
The nuclear recoil contribution can easily by calculated as in
Eq.(\ref{fri8}).
The electron recoil contribution is more complicated. An accurate
treatment of the cross section would
require knowledge of the wavefunctions of all the electrons in the $Na$
and $I$ atoms.
A rough estimate of the electron scattering contribution can be made
by\cite{mmelectron}
considering only the contribution of the loosely
bound (binding energy less than 0.1 keV) outer shell electrons in $Na$
and $I$.  The number
of such loosely bound atomic electrons is 9 for $Na$ and 17 for $I$.
We further approximate these electrons as free and at rest, and compute
the elastic scattering
rate on these electrons. Thus, within this approximation
the cross section has the form given in Eq.(\ref{cs}), with
$\lambda_e = 2\pi\epsilon^2 \alpha^2/m_e$.
The predicted differential interaction rate is then:
\begin{eqnarray}
{dR \over dE_R} &=& 
gN_T  n_{ e'} \int {d\sigma \over dE_R}
{f_{e'}(v) \over k} |v|
d^3v \nonumber \\
&=& g N_T  n_{e'}
{\lambda_e \over E_R^2 } \int^{\infty}_{|v| > v_{min}
(E_R)} {f_{e'}(v) \over k|v|} d^3 v 
\label{55b}
\end{eqnarray}
where $N_T$ is the number of target $NaI$ pairs  per kg of detector and 
$k = (\pi v_0^2 [e'])^{3/2}$ is the Maxwellian distribution
normalization factor.
The quantity $g = 26$, is the number of loosely bound atomic electrons
per $NaI$ pair 
as we discussed above.
Also, $n_{e'}$ is the halo $e'$ number density. Assuming the halo is fully ionized it
is straightforward to show that
\begin{eqnarray}
n_{e'} = \left[ 1 - {\xi_{He'} \over 2} - {\xi_{A'} \over 2} \right] {\rho_{dm} \over m_p} \ .
\end{eqnarray}
Note that the lower velocity limit in Eq.(\ref{55b}),
$v_{min} (E_R)$, 
is given by the kinematic relation:
\begin{eqnarray}
v_{min} &=& \sqrt{ {2E_R\over m_e } } .
\label{v2}
\end{eqnarray}
The velocity integral in Eq.(\ref{55b}) can be analytically solved
leading to:
\begin{eqnarray}
{dR \over dE_R} &=& 
g N_T n_{e'} {\lambda_e \over E_R^2}
\left( {2e^{-x^2}
\over \sqrt{\pi} v_0 [e'] }\right) 
\label{charlie}
\end{eqnarray}
where $x = v_{min}/v_0 [e']$.
Finally, to compare with the experimentally measured rate we
convolve this rate, with a Gaussian [as in Eq.(\ref{fri8})] to take into
account the finite detector resolution.

There are many potential systematic uncertainties in the absolute rate, and we 
consider the following: the uncertainty in the measured
detector resolution [i.e. taking a $2\sigma$ variation of the measured $\sigma_{res}$ given
in Eq.(\ref{26})], 
a $\sim 30\%$ uncertainty in the $e'-e$ scattering cross section,
and a $0.25$ keVee uncertainty in energy calibration.

In figure 7 we give an example of the absolute rate predicted for
the DAMA experiment showing
both electron and nuclear recoil contributions separately, using the aforementioned
systematic uncertainties to minimize the rate. Since the data below 2 keVee 
is formally below the DAMA threshold we do not attempt to fit this data, and 
must await the forthcoming DAMA upgrade which is designed to lower the energy threshold.
The rise in event rate below 2 keVee, which is illustrated in figure 7, 
is a prediction of this model which DAMA can potentially
confirm when they lower their energy threshold.
\vskip 1.0cm
\centerline{\epsfig{file=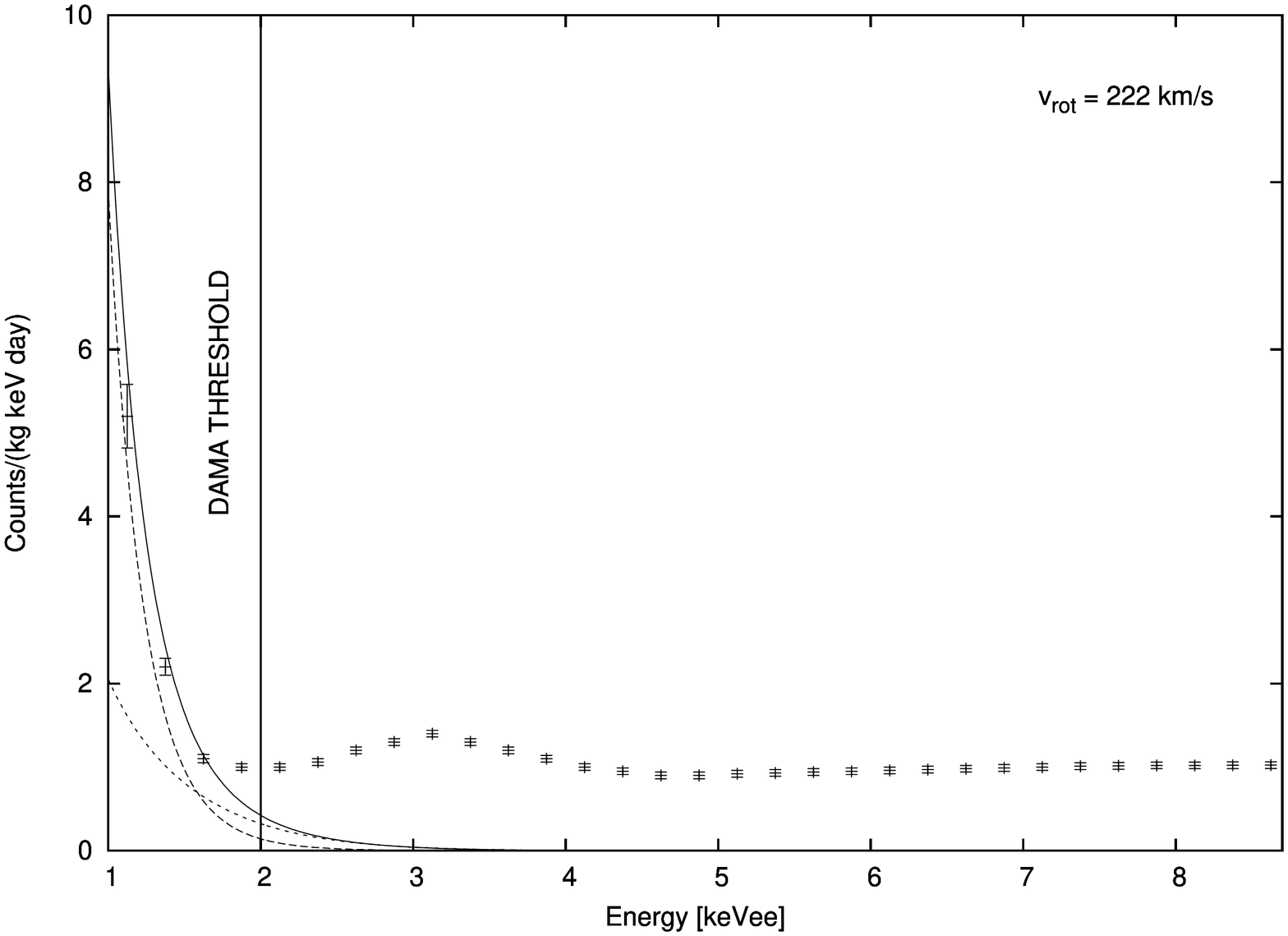,angle=270,width=12.8cm}}
\vskip 0.4cm
\noindent
Figure 7: DAMA absolute rate from nuclear recoils  with negligible channeling (dotted line), 
electron recoils (dashed line) and combined (solid line) for
the parameters: $m_{A'}/m_p = 20$, 
$\epsilon\sqrt{\xi_{A'}} = 5.0 \times 10^{-10}$
and $\epsilon = 1.0 \times 10^{-9}$ ($\Rightarrow \xi_{A'} = 0.25$). 
$v_{rot} = 222$ km/s has been assumed.

\newpage

In figure 8, we given an example of the electron scattering rate predicted for the
CDMS Germanium electron scattering experiment\cite{cdmselectron}, for the same parameters as
chosen for figure 7, together with a simple linear model for the background ($R(background) =
1.9 - 0.09E_R$). 
For the Germanium experiment the resolution is given by\cite{cdmselectron}
\begin{eqnarray}
\sigma = \sqrt{(0.293)^2 + (0.056)^2 E_R/keV }\ keV
\end{eqnarray}
and the rate as in Eq.(\ref{charlie}) but with $g=14$\cite{mmelectron}.
\footnote{
In ref.\cite{mmelectron}, we assumed a cutoff at $E_R = 0.8$ keV. However this is, in fact unjustified,
and here we have no such cutoff except for a phenomenological cutoff at $E_R = 0.2$ keV.
Such a low energy cutoff is
necessary due to the divergence of the cross section in the $E_R \to 0$ limit.}.

\vskip 0.9cm
\centerline{\epsfig{file=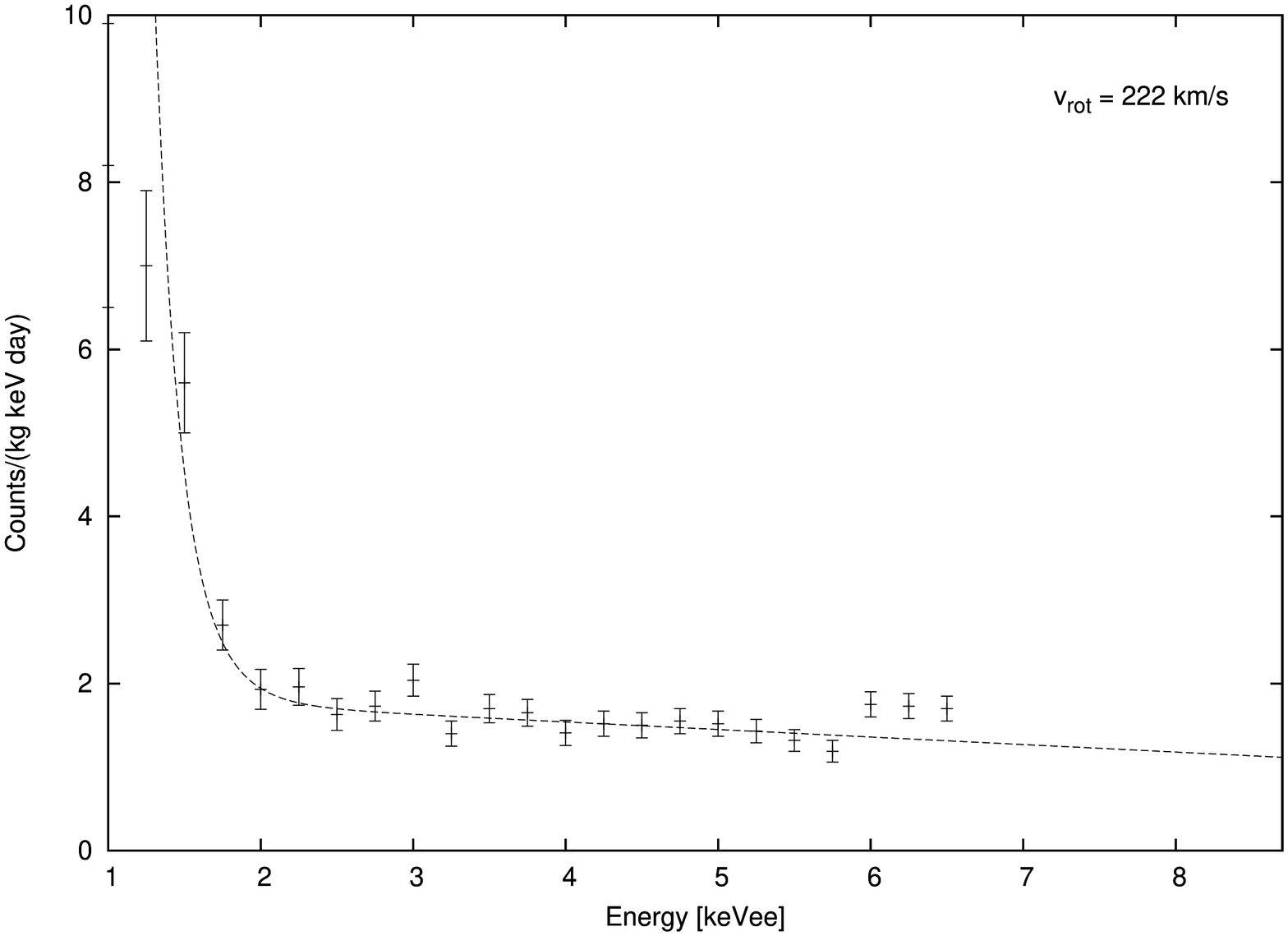,angle=270,width=12.4cm}}
\vskip 0.3cm
\noindent
Figure 8: CDMS/Ge absolute rate from electron recoils for the same parameters as figure 7,
together with a simple linear background model.

\vskip 1cm

The rise in event rate seen below 2 keV in both DAMA and CDMS electron scattering data
is very interesting, but is formally below the threshold of both of these experiments, and
therefore needs to be confirmed by future measurements.
Conservatively, the only limit that we can obtain is by looking at the data above 2 keV.
Demanding that the total rate in the first energy bin above threshold be 
less than 1 cpd/kg/keV,
suggests an upper limit for $\epsilon$. This upper limit depends quite
sensitively on the value of $v_0 [e']$ (and hence also on $v_{rot}$ and 
$\xi_{A'}$). If  $\xi_{A'} \ll 1$, then we
find:
\begin{eqnarray}
%\epsilon \stackrel{<}{\sim} &xxxx&, \ {\rm \ if \ \xi_{A'} \approx 1}
%\nonumber \\
\epsilon \stackrel{<}{\sim} & 2.7 \times 10^{-9} &, \ {\rm \ if \
v_{rot} = 222\ km/s}
\nonumber \\
\epsilon \stackrel{<}{\sim} & 1.5 \times 10^{-9} &, \ {\rm \ if \
v_{rot} = 254\ km/s}
\nonumber \\
\epsilon \stackrel{<}{\sim} & 1.0 \times 10^{-9} &, \ {\rm \ if \
v_{rot} = 286\ km/s}
\end{eqnarray}
If $\xi_{A'} \approx 1$ then we find:
\begin{eqnarray}
\epsilon \stackrel{<}{\sim} & 1.1 \times 10^{-9} &, \ {\rm \ if \
v_{rot} = 222\ km/s}
\nonumber \\
\epsilon \stackrel{<}{\sim} & 0.8 \times 10^{-9} &, \ {\rm \ if \
v_{rot} = 254\ km/s}
\nonumber \\
\epsilon \stackrel{<}{\sim} & 0.6 \times 10^{-9} &, \ {\rm \ if \
v_{rot} = 286\ km/s.}
\label{l65}
\end{eqnarray}
In computing these upper limits we have allowed for some of the
systematic uncertainties by including: 
the uncertainty in the measured
detector resolution [i.e. taking a $2\sigma$ variation of the measured $\sigma_{res}$ given
in Eq.(\ref{26})], 
a $\sim 30\%$ uncertainty in the $e'-e$ scattering cross section,
and a $0.25$ keV uncertainty in energy calibration.
It should be emphasized, though, that the systematic uncertainty can potentially be much
larger in view of the large event rate at low energies which is 
smeared into $E_R \stackrel{>}{\sim}$ 2 keV by the resolution.
In particular the resolution has not been measured at $E_R \stackrel{<}{\sim}$ 2 keVee
and the naive extrapolation might breakdown at low energies. 
It is also possible that the 
resolution might fall off faster than a Gaussian in the tails of the 
distribution, which would weaken
the above limits on $\epsilon$. 
Finally, 
astrophysical uncertainties in modelling the halo will add further
systematic uncertainties 
to the $e'-e$ scattering rate due to its sensitive dependence on $v_0 [e']$.
Departures from spherical symmetry or a rotating halo etc will lead to deviations
from Eq.(\ref{4}), and hence to $v_0 [e']$. [Note though that the 
$A'$ scattering rate is much less sensitive to uncertainties in $v_0$ since
$v_0 [A'] \ll v_{rot}$, and thus such uncertainties will have little
affect on our DAMA/CoGeNT fit].
Thus, given these systematic uncertainties our limits on $\epsilon$ should be 
viewed more as a guide, than
strict upper limits.

Similar limits to the above, with the same caveats 
regarding potentially larger systematic
uncertainties, can be obtained   
from the CDMS electron scattering data.
In combination
with our estimate of $\epsilon \sqrt{\xi_{A'}} \approx (7\pm 3)\times
10^{-10}$ from the DAMA
and CoGeNT experiments, the limits given in Eq.(\ref{l65}) indicate $\xi_{A'} \stackrel{>}{\sim}
10^{-2}$ at $v_{rot} = 222$ km/s, with stronger bounds at higher
$v_{rot}$ values. This suggests
that the mirror sector may have a higher metal content than
the ordinary matter sector.
This is certainly possible, and might be due to a period of rapid mirror
star formation and
evolution during the first few billion years of the Universe (which is
suspected given the
computed high primordial $Y_{He'}\approx 0.9$ abundance\cite{bbn} 
which would dramatically speed up mirror star evolution
by several orders of magnitude\cite{starevolution}). 

\section{CDMS/Ge, EdelweissII and CRESSTII}

The CDMS/Ge and EdelweissII experiments utilize a Germanium target, and both of 
these experiments have found evidence for dark matter interactions. Due to their 
relatively high threshold of $10\ keV_{nr}$ for CDMS/Ge and $20\ keV_{nr}$ for 
Edelweiss, these experiments are not sensitive to the the dominant
$A'$ component. The light mass and narrow velocity dispersion of the $A'$ 
component ensure that the predicted rate for these experiments is much less than 
1 event for their net exposures of approximately 200 kg-days and 322 kg-days 
respectively.  However, these experiments are sensitive to heavier mirror 
dark matter components of the halo, and provide the most sensitive probes of 
the anticipated $Fe'$ component.

In figure 9 we have given an example of the recoil energy spectrum predicted 
for CDMS/Ge (figure 9a) and for EdelweissII (figure 9b). The numerical
work assumed the CDMS/Ge\cite{cdmsge} (EdelweissII\cite{edelweiss}) 
resolution was given by $\sigma_{res} = 0.2$ keV ($\sigma_{res} = 1.0$ keV)
and detection efficiency, $eff = 0.18 + 0.007E_R$ ($eff \simeq 1$).
The value for $v_0 [Fe']$ was obtained from Eq.(\ref{mary2}) assuming that
$\xi_{A'} \ll 1$, i.e.
\begin{eqnarray}
v_0^2 [Fe'] = {v_{rot}^2 \over [m_{Fe}/m_p][2 - \frac{5}{4} \xi_{He'}]}
\end{eqnarray}
where $\xi_{He'} \approx 0.9$.

\vskip 1cm
\centerline{\epsfig{file=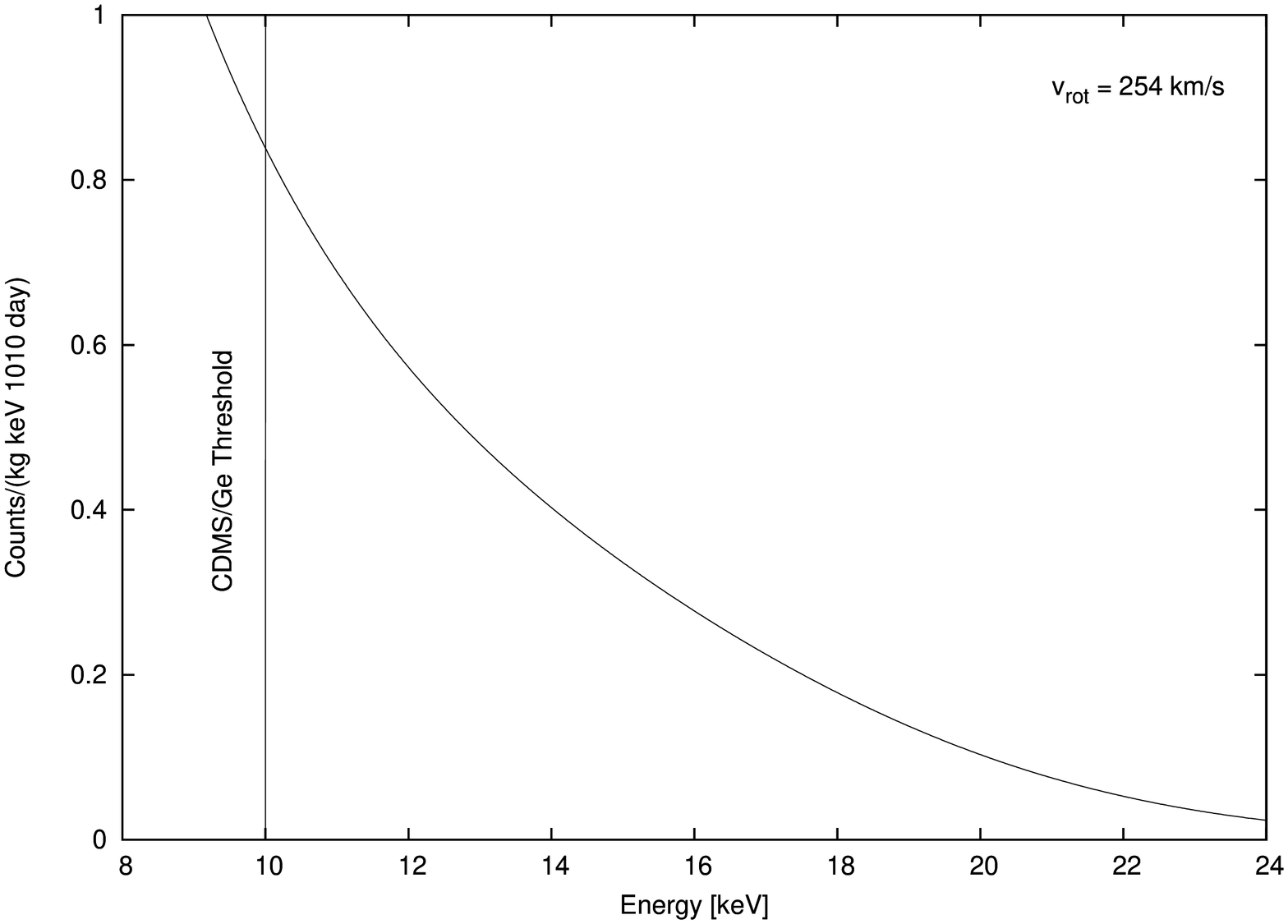,angle=270,width=12.8cm}}
\vskip 0.4cm
\noindent
Figure 9a: 
CDMS/Ge spectrum for $Fe'$ dark matter with 
$\epsilon \sqrt{\xi_{Fe'}} = 5.0 \times 10^{-11}$. We have assumed $v_{rot} = 254$
km/s.
\vskip 1cm
\centerline{\epsfig{file=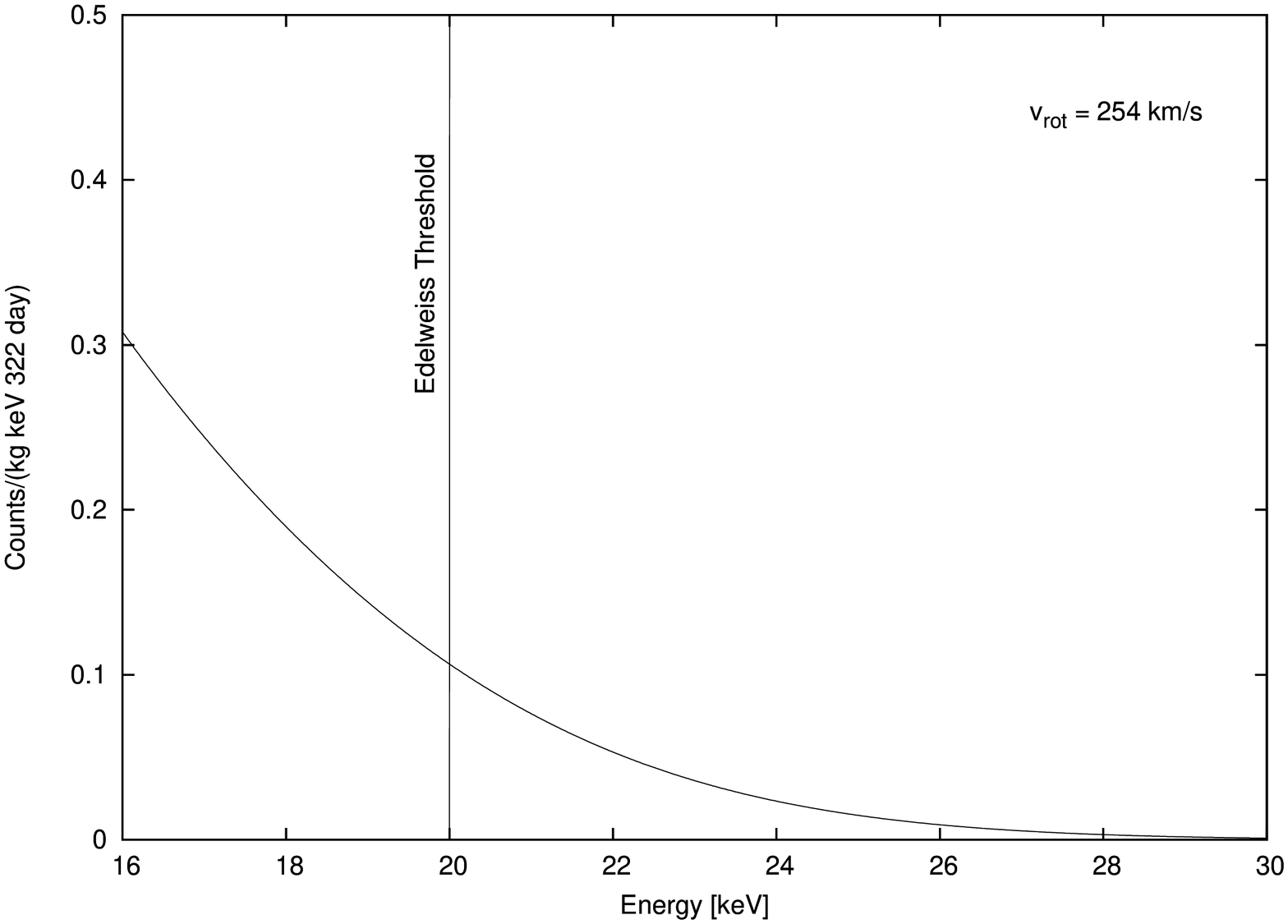,angle=270,width=12.8cm}}
\vskip 0.3cm
\noindent
Figure 9b: 
Edelweiss spectrum for $Fe'$ dark matter with 
$\epsilon \sqrt{\xi_{Fe'}} = 5.0 \times 10^{-11}$. We have assumed $v_{rot} = 254$
km/s.
\vskip 1cm

The CDMS/Ge experiment finds two events at energies 12.3 and 15.5 keV which are clearly 
compatible with the shape of the predicted recoil energy spectrum for CDMS/Ge. 
Edelweiss finds 2 events in their acceptance region with energy 
just above threshold, which is also compatible with 
the shape of the predicted recoil energy spectrum
for EdelweissII. It is therefore plausible that both of these experiments 
have detected $Fe'$ dark matter.
Under this assumption, we can obtain an estimate of 
the parameter: $\epsilon \sqrt{\xi_{Fe'}}$ for each of these experiments:
\begin{eqnarray}
\epsilon \sqrt{Fe'} &=& (3.5^{+2.8}_{-2.0})\times 10^{-11} \ {\rm from\ CDMS/Ge} 
\nonumber \\
\epsilon \sqrt{Fe'} &=& (1.4^{+1.0}_{-0.8})\times 10^{-10} \ 
{\rm from\ EdelweissII} \ .
\end{eqnarray}
Here we have only included the statistical errors at $95\%$ C.L.  
The systematic uncertainties can be quite large, given the rapidly rising 
event rates towards lower $E_R$.  For example, we find that a $20\%$ 
systematic uncertainty in
energy scale would lead to a $\sim 20\%$ uncertainty
in the estimate for $\epsilon \sqrt{\xi_{Fe'}}$ from CDMS/Ge 
and a $50\%$ in the estimate for $\epsilon \sqrt{\xi_{Fe'}}$ from EdelweissII.
Also note that the systematic uncertainties in the form factor begin to be quite significant 
for Edelweiss due to the large recoil energy threshold. 
Clearly systematic uncertainties can reconcile the
two estimates of $\epsilon\sqrt{Fe'}$ from CDMS/Ge and EdelweissII.
The XENON100 experiment, with an anticipated net exposure of over 1000 kg-days should be
able to confirm the presence of a $Fe'$ signal in the near future.

Combining the above estimate for $\epsilon\sqrt{\xi_{Fe'}}$ with our earlier fit 
for $\epsilon \sqrt{\xi_{A'}}$ suggests a $\xi_{Fe'}/\xi_{A'}$ fraction of around 
$\sim 10^{-2}$ which is plausible.  Also note that such a 
small $\xi_{Fe'}$ component does not significantly affect the fit of the 
DAMA and CoGeNT experiments.

The CRESSTII experiment using a $CaWO_4$ target has recently reported 32 
dark matter candidate events, with a background of around 9 events in 
their signal region, which suggests a statistically significant low energy
excess of around 23 events. The threshold of CRESSTII is 10 $keV_{nr}$, 
with the excess of events reported in the oxygen band near threshold.
The CRESSTII experiment is potentially sensitive to both the $A'$ 
component and the $Fe'$ component.  We illustrate this in figure 10, 
where we give the predicted CRESST recoil energy spectrum for an example
with parameters close to the DAMA/CoGeNT best fit.
As this figure illustrates, the $A'$ component is only important
in the region near threshold, while the $Fe'$ contribution is 
somewhat more spread out.  Thus, in principle these two components 
can be distinguished from the observed energy distribution of events (so long
as they both contribute significantly to the signal).
\vskip 1cm
\centerline{\epsfig{file=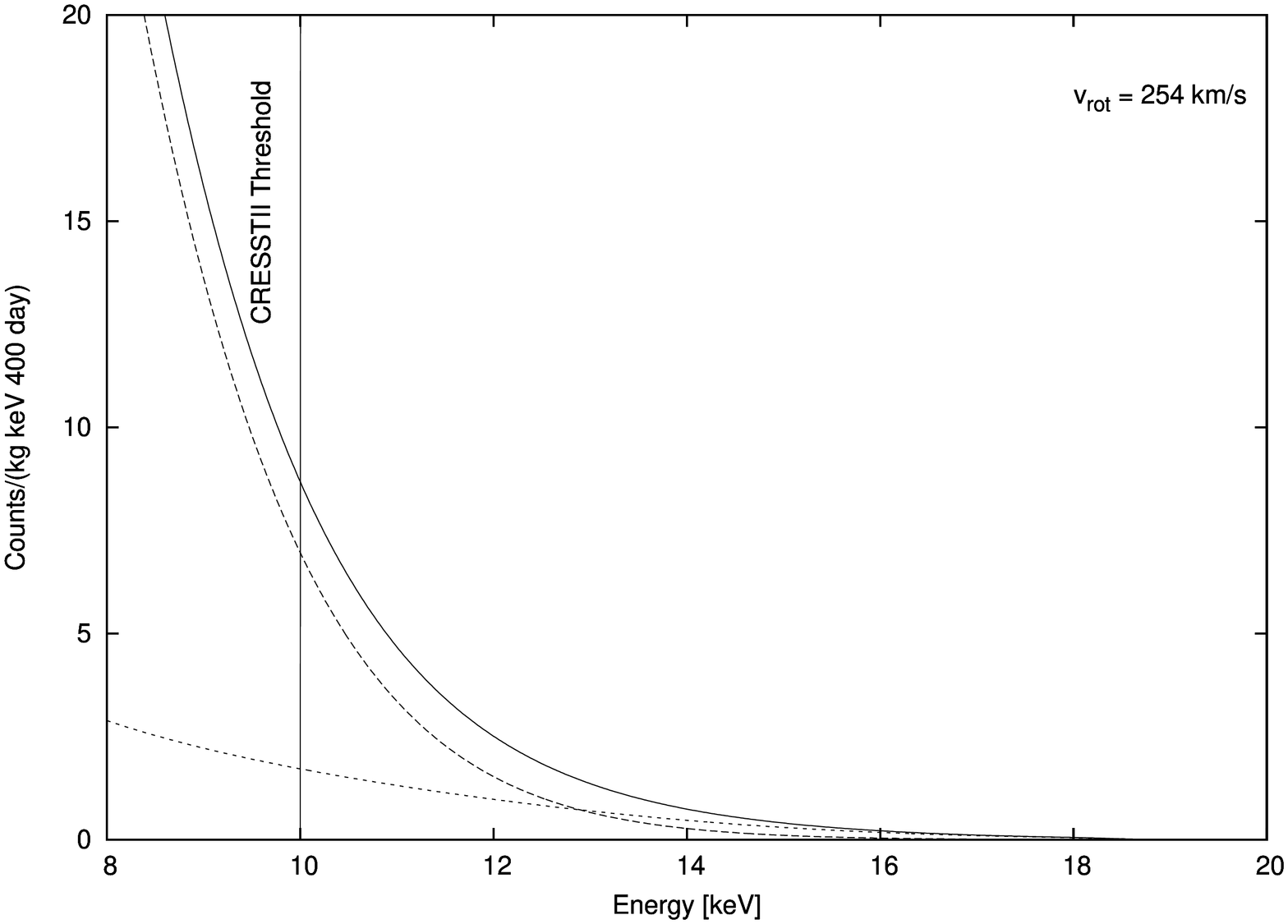,angle=270,width=12.8cm}}
\vskip 0.4cm
\noindent
Figure 10: 
CRESSTII spectrum in the oxygen band for $A', Fe'$ dark matter
assuming $v_{rot} = 254$ km/s.
The dashed line is the $A'$ contribution with parameters 
$\epsilon \sqrt{\xi_{A'}} = 6\times 10^{-10}$, $m_{A'} = 22.0$.
The dotted line is for the $Fe'$ contribution with parameters
$\epsilon \sqrt{\xi_{Fe'}} = 1.0 \times 10^{-10}$. 
The solid line is the sum of the two contributions.
[100\% detection efficiency has been assumed, and a resolution of $0.3$ keV].
\vskip 1cm

We have found that the $A'$ contribution has a very large 
annual modulation in the energy region above threshold. In fact,  
\begin{eqnarray}
{\overline{dR}^1 \over dE_R^m } \approx 0.5 {\overline{dR}^0\over dE_R^m} \ .
\end{eqnarray}
The large annual modulation results because the interactions arise 
from $A'$ particles in the tail of 
the narrow Maxwellian distribution and therefore can be greatly affected by 
small changes in the velocity of the Earth. Thus, if $A'$ does
contribute significantly to the signal, an examination of events in 
the $Ethreshold < E < Ethreshold+3 \ keV$ bin should show a statistically 
significant annual modulation with just 1-2 years of data. The annual modulation 
predicted for the $Fe'$ events is smaller, but still quite large:
\begin{eqnarray}
{\overline{dR}^1 \over dE_R^m } \approx 0.16 {\overline{dR}^0\over dE_R^m} \ 
\end{eqnarray}
and could also be eventually seen provided that a significant proportion
of the events are due to $Fe'$ interactions.

\section{conclusion}

In conclusion, we have confronted
the mirror dark matter theory with the most recent experimental data. 
We examined the DAMA 
experiments allowing for the possibility of a negligible channeling
fraction and showed
that under that assumption the impressive DAMA signal can be fully
explained. 
Mirror dark matter can simultaneously explain
the CoGeNT low energy excess, and remains compatible with
the results of the other experiments, including interesting hints of
dark matter detection from
CDMS/Ge, EdelweissII and CRESSTII.
Taking into account some of the possible systematic uncertainties
in quenching factor, detector resolution, galactic rotational velocity
and velocity dispersion, we have mapped out the allowed regions of parameter
space
in the $m_{A'}, \epsilon\sqrt{\xi_{A'}}$ plane [Figures 4,5].
The net result is that the 
mirror dark matter candidate can explain all of the existing direct
detection
experiments, with parameters $\epsilon \sqrt{\xi_{A'}} = (7 \pm 3)
\times 10^{-10}$,
$m_{A'}/m_p = 22\pm 8$, $\xi_{A'} \stackrel{>}{\sim} 10^{-2}, 
\ \xi_{Fe'}/\xi_{A'} \sim 10^{-2}$.

Importantly this theory will soon be more stringently tested by:
a) Further data from DAMA: in the near future the DAMA collaboration
plan to upgrade their
experiment with the aim of lowering their energy threshold. As illustrated
in figure 1, they 
should see a change in sign of their modulation amplitude between
1.0-2.0 keVee.
b) The CRESSTII experiment, with light target element $O'$ and threshold $10$ keV
is potentially sensitive to both $A'$ and $Fe'$ components. They should find a rapidly 
falling energy spectrum
with most of their events between 10 and 14 keV, with a very large annual modulation.
c) The CDMS/Si experiment is sensitive to the dominant mirror metal component, $A'$. 
This experiment currently provides the strongest constraint from
the null experiments and limits $m_{A'}/m_p \stackrel{<}{\sim}$ 30. Further data from CDMS/Si
should either see a signal, or produce tighter constraints on $m_{A'}$.
d) More data from the CoGeNT experiment should be helpful. Of particular note, is that mirror dark matter
predicts a detectable annual modulation signal for CoGeNT (figure 6) with the distinctive feature that
it changes in sign at energies around $E_R \approx 0.5-0.8$ keVee. 
%A negative annual modulation amplitude means that CoGeNT should see {\it more} events 
%during the (northern) winter/fall than the summer/spring.
e) Very sensitive but typically higher threshold experiments, such as
CDMS/Ge, EdelweissII and XENON100
can potentially probe the heavier $\sim Fe'$
component which should exist at some level. The two events seen in
CDMS/Ge and in EdelweissII are consistent
with this component, and suggests $\xi_{Fe'}/\xi_{A'} \sim 10^{-2}$.
f) In the longer term, directional experiments will be important
due to the low velocity dispersion of mirror dark matter [$v_0[A'] \ll v_{rot}$]. 
\footnote{A dark matter
detection experiment in the southern hemisphere would be very useful because, due to the Earth's orientation, 
it could sensitively probe the possible diurnal variation if there are
accumulated mirror particles in the Earth's core. }

Mirror dark matter is not expected to show up in collider experiments
through the photon-mirror photon
kinetic mixing induced interactions. However, the Higgs - mirror Higgs
quartic interaction [Eq.(\ref{mix})]
leads to modifications of the properties of the Higgs boson\cite{flv}
which can potentially be
observed at the LHC and Tevatron\cite{rayig}.  Sensitive
orthopositronium 
studies\cite{glashow}, might be able to directly
probe $\epsilon \sim 10^{-9}$ and thus provide further tests of the
mirror dark matter scenario.
Such an experiment has been proposed recently in ref.\cite{ortho}.

\vskip 1cm
\noindent 
{\bf Acknowledgments}
\vskip 0.2cm
\noindent
This work was supported by the Australian Research Council.


\begin{thebibliography}{999}

\bibitem{dama}
R. Bernabei {\it et al}. (DAMA Collaboration), 
Riv. Nuovo Cimento. 26, 1 (2003) [astro-ph/0307403]; Int. J. Mod.
Phys. D13, 2127 (2004); Phys. Lett. B480, 23 (2000).

\bibitem{dama2}
R. Bernabei {\it et al}. (DAMA Collaboration), 
Eur. Phys. J. C56: 333 (2008) [arXiv:0804.2741].

\bibitem{dama3}
R. Bernabei {\it et al}. (DAMA Collaboration), arXiv: 1002.1028.

\bibitem{cogent}
C. E. Aalseth {\it et al.} (CoGeNT Collaboration),  arXiv:1002.4703.

\bibitem{cdmsge}
Z. Ahmed {\it et al.} (CDMS Collaboration), arXiv: 0912.3592.

\bibitem{cdmselectron}
Z. Ahmed {\it et al.} (CDMS Collaboration), Phys. Rev. D81: 042002 (2010)
[arXiv: 0907.1438].

\bibitem{cresst}
W. Seidel (for the CRESST collaboration), Talk given at IDM2010, July 2010.

\bibitem{edelweiss}
E. Armengaud (for the Edelweiss Collaboration), Talk given at IDM2010, July 2010;
Phys. Lett. B687, 294 (2010) [arXiv:0912.0805].

\bibitem{mm}
R. Foot, Phys. Rev. D69, 036001 (2004) [hep-ph/0308254];
Mod. Phys. Lett. A19,
1841 (2004) [astro-ph/0405362]; Phys. Rev. D74, 023514 (2006)
[astro-ph/0510705].

\bibitem{mm2}
R. Foot, Phys. Rev. D78, 043529 (2008) [arXiv: 0804.4518].

\bibitem{mmelectron}
R. Foot, Phys. Rev. D80, 091701 (2009) [arXiv:0909.3126].

\bibitem{mmcdms}
R. Foot, Phys. Rev. D81, 087302 (2010) [arXiv:1001.0096].

\bibitem{mmcogent}
R. Foot, arXiv:1004.1424.

\bibitem{flv}
R. Foot, H. Lew and R. R. Volkas, Phys. Lett. B272, 67 (1991);
Mod. Phys. Lett. A7, 2567 (1992).

\bibitem{review}
R. Foot, Int. J. Mod. Phys. D13, 2161 (2004)
[astro-ph/0407623].

\bibitem{sph}
R. Foot and R. R. Volkas,
Phys. Rev. D70, 123508 (2004) [astro-ph/0407522].

\bibitem{he}
R. Foot and X-G. He, Phys. Lett. B267, 509 (1991).

\bibitem{lee}
T. D. Lee and C. N. Yang, Phys. Rev. 104, 256 (1956);
I. Kobzarev, L. Okun and I. Pommeranchuk, Sov. J. Nucl. Phys. 3, 837
(1966);
M. Pavsic, Int. J. Theor. Phys. 9, 229 (1974).

\bibitem{macholimits}
C. Alcock {\it et al.} (MACHO collaboration), ApJ, 542, 281 (2000)
[arXiv: astro-ph/0001272];
P. Tisserand {\it et al.}, (EROS Collaboration) Astron. Astrophys, 469, 387
(2007) [arXiv: astro-ph/0607207].

\bibitem{bulletcluster}
D. Clowe {\it et al.}, Astrophys. J. 648, L109 (2006)
[astro-ph/0608407].

\bibitem{silagadze}
Z. K. Silagadze, arXiv:0808.2595.

\bibitem{lab}
R. Foot, A. Yu. Ignatiev and R. R. Volkas, Phys. Lett. B503, 355 (2001)
[arXiv: astro-ph/0011156];
R. Foot, Int. J. Mod. Phys. A19 3807 (2004) [astro-ph/0309330];
R. Foot and Z. K. Silagadze, Int. J. Mod. Phys. D14, 143 (2005)
[astro-ph/0404515];
P. Ciarcelluti and R. Foot, 
Phys. Lett. B679, 278 (2009) [arXiv: 0809.4438].
See also, S. Davidson, S. Hannestad and G. Raffelt, JHEP 5, 3
(2000) [arXiv: hep-ph/0001179].

\bibitem{rafelt}
G. Raffelt, {\it Stars as Laboratories for Fundamental Physics,}
Chicago University Press (1996).

\bibitem{bbn}
P. Ciarcelluti and R. Foot, Phys. Lett. B690, 462 (2010)
[arXiv:1003.0880].

\bibitem{starevolution}
Z. Berezhiani {\it et al.}, Astropart. Phys. 24, 495 (2006)
[astro-ph/0507153].

\bibitem{some}
Z. Berezhiani, D. Comelli and F. L. Villante,
Phys. Lett. B503, 362 (2001) [hep-ph/0008105];
L. Bento and Z. Berezhiani, Phys. Rev. Lett. 87, 231304 (2001)
[hep-ph/0107281]; 
A. Yu. Ignatiev and R. R. Volkas, Phys. Rev. D68, 023518 (2003)
[hep-ph/0304260];
R. Foot and R. R. Volkas, Phys. Rev. D68, 021304 (2003)
[hep-ph/0304261]; Phys. Rev. D69, 123510 (2004) [hep-ph/0402267];
Z. Berezhiani, P. Ciarcelluti, D. Comelli and F. L. Villante,
Int. J. Mod. Phys. D14, 107 (2005) [astro-ph/0312605];
P. Ciarcelluti, Int. J. Mod. Phys. D14, 187 (2005) [astro-ph/0409630];
Int. J. Mod. Phys. D14, 223 (2005) [astro-ph/0409633].

\bibitem{helm}
R. H, Helm, Phys. Rev. 104, 1466 (1956).

\bibitem{smith}
J. D. Lewin and P. F. Smith, Astropart. Phys. 6, 87 (1996).

\bibitem{rot}
M. J. Reid {\it et al.}, Astrophys. J. 700, 137 (2009) [arXiv:
0902.3913].


\bibitem{idea}
A. K. Drukier, K. Freese and D. N. Spergel, Phys. Rev. D33, 3495 (1986);
K. Freese, J. A. Frieman and A. Gould, Phys. Rev. D37, 3388 (1988).

\bibitem{damares}
R. Bernabei {\it et al}. (DAMA Collaboration),
Nucl. Instrum. Meth. A592: 297 (2008) [arXiv: 0804.2738].


\bibitem{gelmini}
N. Bozorgnia, G.B. Gelmini and P. Gondolo, arXiv: 1006.3110.


\bibitem{damachan}
R. Bernabei {\it et al}. (DAMA Collaboration), Eur. Phys. J. C53, 205
(2008) [arXiv: 0710.0288].

\bibitem{cogent2}
D. Hooper, J. I. Collar, J. Hall and D. McKinsey, arXiv: 1007.1005.

\bibitem{cog2}
C. E. Aalseth {\it et al.} (CoGeNT Collaboration),  Phys. Rev. Lett.
101, 251301 (2008) [arXiv: 0807.0879].

\bibitem{cdmssi}
J. P. Filippini, Ph.D thesis, 2008.

\bibitem{xenon100}
E. Aprile {\it et al.} (XENON100 Collaboration), arXiv: 1005.0380
(2010).

\bibitem{collar}
J. I. Collar and D. N. McKinsey, arXiv:1005.0838; arXiv: 1005.3723;
The Xenon Collaboration, arXiv: 1005.2615;
J. I. Collar, arXiv: 1006.2031; P. Sorensen, arXiv: 1007.3549.


\bibitem{rayig}
A. Yu. Ignatiev and R. R. Volkas,
Phys. Lett. B487, 294 (2000) [hep-ph/0005238];
R. Barbieri, T. Gregoire and L. J. Hall, hep-ph/0509242;
W. Li, P. Yin and S. Zhu, Phys. Rev. D76, 095012 (2007) [arXiv:
0709.1586].

\bibitem{glashow}
S. L. Glashow, Phys. Lett. B167, 35 (1986); R. Foot and S. N. Gninenko, 
Phys. Lett. B480, 171 (2000) [hep-ph/0003278].


\bibitem{ortho}
P. Crivelli {\it et al.}, arXiv: 1005.4802.

% hereeeeeeeeeeee xxxxxxxxxxxxxxxxxxxxxxxxxxxxxxxxx





%\bibitem{xenon}
%J. Angle {\it et al}. (XENON Collaboration), Phys. Rev. Lett. 100,
%021303 (2008) [arXiv:0706.0039].
%

%\bibitem{talk}
%See e.g. R. Foot, arXiv: 0907.0048 (2009).



%\bibitem{xenon2}
%A. Manzur {\it et al.}, arXiv: 0909.1063 (2009);
%E. Aprile {\it et al.} arXiv: 0810.0274 (2008).



\end{thebibliography}
\end{document}